\documentclass[runningheads]{llncs}
\usepackage{graphicx}
\usepackage{newlfont}
\usepackage{amsmath}
\usepackage{graphicx}
\usepackage{amssymb}
\usepackage{latexsym}
\usepackage{stmaryrd}
\usepackage{braket}
\usepackage{enumerate}
\usepackage{booktabs}
\usepackage{mathtools}
\usepackage{array}
\usepackage{paralist}
\usepackage{hyperref}
 
\usepackage{color}

\DeclareMathOperator{\sat}{sat}
\DeclareMathOperator{\solvdeg}{solv.deg}
\DeclareMathOperator{\maxGB}{max.GB.deg}
\DeclareMathOperator{\ini}{in}

\DeclareMathOperator{\reg}{reg}

\DeclareMathOperator{\height}{height}
\DeclareMathOperator{\supp}{supp}
\DeclareMathOperator{\Syz}{Syz}
\DeclareMathOperator{\Triv}{Triv}

\newcommand{\kk}{k}            
\newcommand{\FF}{\mathbb{F}}  
\newcommand{\dregF}[1]{d_{\mathrm{reg}}(#1)}
\newcommand{\dregD}[1]{\delta_{\mathrm{reg}}(#1)}
\newcommand{\ireg}[1]{i_{\mathrm{reg}}(#1)}
\newcommand{\PP}{\mathbb{P}}

\begin{document}
\title{Solving multivariate polynomial systems and an invariant from commutative algebra}
\titlerunning{Solving multivariate polynomial systems}
%
\author{Alessio Caminata\inst{1}\orcidID{0000-0001-5227-807X} \and
Elisa Gorla\inst{2}\thanks{Corresponding author.}}
\authorrunning{A.~Caminata and E.~Gorla}
%
\institute{Dipartimento di Matematica, Universit\`a degli Studi di Genova, \\via Dodecaneso 35, 16146 Genova, Italy\\
\email{caminata@dima.unige.it}\\
\and
Institut de Math\'{e}matiques, Universit\'{e} de Neuch\^{a}tel\\Rue Emile-Argand 11, CH-2000 Neuch\^{a}tel, Switzerland\\
\email{elisa.gorla@unine.ch}}
\maketitle              
\begin{abstract}
The complexity of computing the solutions of a system of multivariate polynomial equations by means of Gr\"obner bases computations is upper bounded by a function of the solving degree.
In this paper, we discuss how to rigorously estimate the solving degree of a system, focusing on systems arising within public-key cryptography. In particular, we show that it is upper bounded by, and often equal to, the Castelnuovo-Mumford regularity of the ideal generated by the homogenization of the equations of the system, or by the equations themselves in case they are homogeneous. We discuss the underlying commutative algebra and clarify under which assumptions the commonly used results hold. In particular, we discuss the assumption of being in generic coordinates (often required for bounds obtained following this type of approach) and prove that systems that contain the field equations or their fake Weil descent are in generic coordinates. We also compare the notion of solving degree with that of degree of regularity, which is commonly used in the literature. We complement the paper with some examples of bounds obtained following the strategy that we describe.

\end{abstract}
%
%
%

\section*{Introduction}\label{section-introduction}

Polynomial system solving plays an important role in many areas of mathematics. In this paper, we discuss how to solve a system of multivariate polynomial equations by means of Gr\"obner bases techniques and estimate the complexity of polynomial system solving. Our motivation comes from public-key cryptography, where the computational problem of solving polynomial systems of equations plays a major role.

In multivariate cryptography, the security relies on the computational hardness of finding the solutions of a system of polynomial equations over a finite field. One can use similar strategies in order to produce public-key encryption schemes and digital signature algorithms, whose security relies on this problem. For signature schemes, e.g., the public key takes the form of a polynomial map
$$\begin{array}{rcl} \mathcal{P}:\FF_q^n & \longrightarrow & \FF_q^r \\
(a_1,\ldots,a_n) & \longmapsto & (f_1(a_1,\ldots,a_n),\ldots,f_r(a_1,\ldots,a_n))
\end{array}$$ 
where $f_1,\dots,f_r\in\FF_q[x_1,\ldots,x_n]$ are multivariate polynomials with coefficients in a finite field $\FF_q$. The secret key allows Alice to easily invert the system $\mathcal{P}$. In order to sign the hash $b$ of a message, Alice computes $a\in\mathcal{P}^{-1}(b)$ and sends it to Bob. Bob can readily verify the validity of the signature by checking whether $\mathcal{P}(a)=b$. An illegitimate user Eve who wants to produce a valid signature without knowing Alice's secret key is faced with the problem of solving the polynomial system of $r$ equations in $n$ variables
$$\left\{\begin{array}{c}
f_1(x_1,\ldots,x_n)=b_1\\
\vdots \\
f_r(x_1,\ldots,x_n)=b_r
\end{array}\right.$$
Even without knowing Alice's secret key, Eve may be able to exploit the structure of $\mathcal{P}$ in order to solve the system. Such an approach is largely used and the adopted strategies vary significantly from one cryptographic scheme to another. Moreover a direct attack is always possible, i.e., Eve may try to solve the system by computing a Gr\"obner basis of it.
Therefore, being able to estimate the computational complexity of solving a multivariate polynomial system gives an upper bound of the security of the corresponding cryptographic scheme, and is therefore highly relevant. In this context, the complexity of solving a polynomial system is typically large enough to make the computation unfeasible, since being able to compute a solution would enable the attacker to forge a digital signature or to decrypt an encrypted message. We emphasize that the security of multivariate cryptographic schemes is a theme of high current interest. For example, the National Institute of Standards (NIST) is in the process of selecting post-quantum cryptographic schemes for standardization. Three digital signature algorithms were selected as finalists in Round 3 by NIST in July 2020~\cite{NIST3}, one of which is a multivariate scheme.

Multivariate polynomial systems also appear in connection with the Discrete Logarithm Problem (DLP) on an elliptic or hyperelliptic curve. An index calculus algorithm for solving the DLP on an abelian variety was proposed in~\cite{Gau09}. The relation-collection phase of the algorithm relies on Gr\"obner bases computations to solve a large number of polynomial systems. These systems usually do not have any solutions, but, whenever they have one, they produce a decomposition of a point of the abelian variety over the chosen factor base. In contrast with polynomial systems arising within multivariate cryptography, it is feasible to solve the polynomial systems arising within index calculus algorithms. Nevertheless, it is important to be able to accurately estimate the complexity of solving them. In fact, the complexity of solving these systems has a direct impact on the complexity of the corresponding index calculus algorithm to solve the DLP.

Estimating the complexity of solving multivariate polynomial systems is relevant within public-key cryptography. In this context, we usually wish to compute the solutions over a finite field of a system of multivariate polynomial equations. Typically, the systems have one, or few, or no solutions, not only over the chosen finite field, but also over its algebraic closure. Moreover, the equations are usually not homogeneous. The degrees of the equations are often small for systems coming from multivariate cryptography, but they can be large for systems arising within index calculus algorithms. Similarly, the number of equations and of variables can vary. Therefore, in this paper we concentrate on finite fields and on non homogeneous systems, which have a finite number of solutions over the algebraic closure. We however do not make assumptions on the number of variables, the number of equations and their degrees.

This paper is devoted to an in-depth discussion of how to estimate the complexity of computing a Gr\"obner basis for a system of multivariate polynomial equations. As said before, our focus is on finite fields and on systems that have a finite number of solutions over the algebraic closure. At the same time, we try to keep the discussion more general, whenever possible. We often concentrate on systems which are not homogeneous, not only because this is the relevant case for cryptographic applications, but also because it is the most difficult case to treat.

After recalling in Section~\ref{section-preliminaries} the commutative algebras preliminaries that will be needed throughout the paper, in Section~\ref{section-shapelemma} we discuss in detail the relation between computing Gr\"obner bases and solving polynomial systems. This connection is often taken for granted within the cryptographic community, as are the necessary technical assumptions. In Section~\ref{section-shapelemma} we discuss in detail what these technical assumptions are and what can be done when they are not satisfied. We also show in Theorem~\ref{polytimeequiv} that, under the usual assumptions, solving a polynomial system of equations is polynomial-time-equivalent to computing a Gr\"obner basis of it. We conclude with Subsection~\ref{subs:fieldeqn}, where we discuss the feasibility of adding the field equations to a system. 

Section~\ref{section-solvingdegree} is the core of the paper. After establishing the setup that we will be adopting, we prove some results on Gr\"obner bases and homogenization/dehomogenization. They allow us to compare, in Theorem~\ref{cor-solvdegIh}, the solving degree of a system, the solving degree of its homogenization, and the solving degree of the homogenization of the ideal generated by its equations.
Combining these results with a classical theorem by Bayer and Stillman~\cite{BS87}, we obtain Theorem~\ref{theorem-regularity=solvingdegree-homog} and Theorem~\ref{theorem-regularity=solvingdegree}, where we show that the Castelnuovo-Mumford regularity upper bounds the solving degree of a system, and recover Macaulay's Bound in Corollary~\ref{corollary-solvdegzerodimensional}. These results hold under the assumption that the homogenized system of equations is in generic coordinates, an assumption that is often overlooked in the cryptographic literature and that we discuss in Section~\ref{section-preliminaries}. In Theorem~\ref{thm-solvdegingencoord} we prove that any system that contains the field equations or their fake Weil descent is in generic coordinates. 

In Section~\ref{section-dreg} we discuss the relation between solving degree and degree of regularity. The latter concept is commonly used in the cryptographic literature and often used as a proxy for the solving degree. In Section~\ref{section-dreg} we discuss the limitations of this approach. In particular, Example~\ref{ex-tzs} and  Example~\ref{ex-itopnonzerodim} are examples of systems coming from index calculus for which, respectively, the degree of regularity is strictly smaller than the solving degree and the degree of regularity is not defined.

Finally, Section~\ref{section-minrank} is meant as an example of how the results from Section~\ref{section-solvingdegree}, in combination with known commutative algebra results, easily provide estimates for the solving degree. In particular, Theorem~\ref{thm-en} and Theorem~\ref{CSideals} give bounds for the solving degree of polynomial systems coming from the MinRank Problem.

{\bf Acknowledgements:} The authors are grateful to Albrecht Petzoldt for help with MAGMA computations, to Wouter Castryck and Sara Gharahbeigi for pointing out imprecisions in earlier versions of this paper, and to Marc Chardin, Teo Mora, Christophe Petit, and Pierre-Jean Spaenlehauer for useful discussions on the material of this paper. This work was made possible thanks to funding from Armasuisse.

\section{Preliminaries}\label{section-preliminaries}

\par In this section we introduce the basic notations and terminology from commutative algebra that we need in the rest of the paper.
All the definitions and the proofs of the results that we quote here are extensively covered in the books \cite{KR00}, \cite{KR05}, \cite{KR16}, and \cite{CLO07}.

\subsection{Polynomial rings and term orders}
\par We work in a polynomial ring $R=\kk[x_1,\dots,x_n]$ in $n$ variables over a field $\kk$.
An element $f\in R$ is a polynomial, and may be written as a finite sum $f=\sum_{\nu} a_{\nu}x^{\nu}$, where $\nu\in\mathbb{N}^n$, $a_{\nu}\in\kk$, and $x^{\nu}=x_1^{\nu_1}\cdots x_n^{\nu_n}$.
A polynomial of the form $a_{\nu}x^{\nu}$ is called a monomial of degree $|\nu|=\nu_1+\cdots+\nu_n$. In particular, every polynomial $f$ is a sum of monomials. 
The degree of $f$, denoted by $\deg(f)$, is the maximum of the degrees of the monomials appearing in $f$.
If all these monomials have the same degree, say $d$, then $f$ is \emph{homogeneous} of degree $d$.
A monomial $a_{\nu}x^{\nu}$ with $a_{\nu}=1$ is \emph{monic}. A monic monomial is also called a \emph{term}.\\

\vspace{1mm}
\textbf{Notation. }
Given a system of polynomials $\mathcal{F}=\{f_1,\dots,f_r\}\subseteq R$ we denote by $(\mathcal{F})=(f_1,\dots,f_r)$ the ideal that they generate, that is $(f_1,\dots,f_r)=\{\sum_{i=1}^rp_if_i: \ p_i\in R\}$. 
\vspace{4mm}
\par The list $\mathcal{F}=\{f_1,\dots,f_r\}$ is called a system of generators of the ideal $I=(\mathcal{F})$. $\mathcal{F}$ is a \emph{minimal system of generators} for $I$ if the ideal generated by any non-empty proper subset of $\mathcal{F}$ is strictly contained in $I$. If the polynomials $f_1,\dots,f_r$ are homogeneous, then we say that the system $\mathcal{F}$and the ideal $I$ are \emph{homogeneous}.

\begin{remark}
Let $I$ be an ideal of $R$ minimally generated by homogeneous polynomials $f_1,\dots,f_r$. Then every homogeneous minimal system of generators of $I$ consists of $r$ polynomials of the same degrees as $f_1,\dots,f_r$. 
\end{remark}

For any degree $d\in\mathbb{Z}_+$, denote by $R_d$ the $d$-th homogeneous component of $R$. $R_d$ is generated as a $\kk$-vector space by the monomials of $R$ of degree $d$. If $I\subseteq R$ is homogeneous, we let $I_d=I\cap R_d$ be the $\kk$-vector space of homogenous polynomials of degree $d$ in $I$.

\medskip 

\par We denote by $\mathbb{T}$ the set of terms of $R$.
A \emph{term order} on $R$ is a total order $\tau$ on the set $\mathbb{T}$, which satisfies the following additional properties:
\begin{compactenum}
\item $m\leq_{\tau} n$ implies $p\cdot m\leq_{\tau} p\cdot n$ for all $p,m,n\in\mathbb{T}$;
\item $1\leq_{\tau} m$ for all $m\in\mathbb{T}$.
\end{compactenum}
If in addition $m<_{\tau} n$ whenever $\deg(m)<\deg(n)$, we say that the term order $\tau$ is \emph{degree-compatible}.

\begin{example}[Lexicographic order]
Let $x^{\alpha}$ and $x^{\beta}$ be two terms in $R$.
We say that  $x^{\alpha}>_{LEX} x^{\beta}$ if the leftmost non-zero entry in the vector $\alpha-\beta\in\mathbb{Z}^n$ is positive.
This term order is called \emph{lexicographic} and it is not degree-compatible. We denote it by $LEX$.
\end{example}

\begin{example}[Degree reverse lexicographic order]
Let $x^{\alpha}$ and $x^{\beta}$ be two terms in $R$.
We say that  $x^{\alpha}>_{DRL} x^{\beta}$ if $|\alpha|>|\beta|$, or $|\alpha|=|\beta|$ and the rightmost non-zero entry in $\alpha-\beta\in\mathbb{Z}^n$ is negative.
This term order is called \emph{degree reverse lexicographic} ($DRL$ for short) and it is degree-compatible.
\end{example}

\par Let $f=\sum_{i\in \mathcal{I}}a_{i}m_i\in R\setminus\{0\}$ be a polynomial, where $a_i\in\kk\setminus\{0\}$, and $m_i\in\mathbb{T}$ are distinct terms.
We fix a term order $\tau$ on $R$.
The \emph{initial term} or \emph{leading term} of $f$ with respect to $\tau$ is the largest term appearing in $f$, that is $\ini_{\tau}(f)=m_j$, where $m_j>m_i$ for all $i\in \mathcal{I}\setminus\{j\}$.
The \emph{support} of $f$ is $\supp(f)=\{m_i: \ i\in \mathcal{I}\}$.
Given an ideal $I$ of $R$, the \emph{initial ideal} of $I$ is 
\begin{equation*}
\ini_{\tau}(I)=(\ini_{\tau}(f): \ f\in I\setminus\{0\}).
\end{equation*}
\begin{definition}
Let $I$ be an ideal of $R$. A set of polynomials $\mathcal{G}\subseteq{I}$ is a \emph{Gr\"obner basis} of $I$ with respect to $\tau$ if 
$\ini_{\tau}(I)=(\ini_{\tau}(g): \ g\in\mathcal{G})$.
A Gr\"obner basis is \emph{reduced} if $m\not\in (\ini_{\tau}(h): \ h\in\mathcal{G}\setminus\{g\})$ for all $g\in\mathcal{G}$ and $m\in\supp(g)$.
\end{definition}

\medskip

Sometimes we will need to consider a field extension. 
At the level of the ideal, this corresponds to looking at the ideal generated by the equations in a polynomial ring over the desired field extension.

\begin{definition}\label{def-extension}
Let $I=(f_1,\ldots,f_r)\subseteq R=\kk[x_1,\dots,x_n]$, let $K\supseteq\kk$ be a field extension. We denote by $IK[x_1,\dots,x_n]$ the \emph{extension} of $I$ to $K[x_1,\dots,x_n]$, i.e. the ideal of $K[x_1,\dots,x_n]$ generated by $f_1,\ldots,f_r$. In symbols, $IK[x_1,\dots,x_n]=(f_1,\ldots,f_r)\subseteq K[x_1,\dots,x_n]$.
\end{definition}

\subsection{Zero loci of ideals}

\par We are mostly interested in ideals, whose zero locus is finite. 

\begin{definition}\label{def-zerodim} 
The \emph{affine zero locus} of an ideal $I=(f_1,\ldots,f_r)\subseteq R$ over the algebraic closure $\bar{\kk}$ of $\kk$ is $$\mathcal{Z}(I)=\{P\in\bar{\kk}^n:\ f(P)=0 \ \mbox{ for all } f\in I\}=\{P\in\bar{\kk}^n:\ f_1(P)=\ldots=f_r(P)=0\}.$$ We also denote it by $\mathcal{Z}(f_1,\ldots,f_r)$.
\end{definition}

\begin{definition}\label{def-zerodim-homog}
The \emph{projective zero locus} of a homogeneous ideal $I=(f_1,\ldots,f_r)\subseteq R$ over the algebraic closure $\bar{\kk}$ of $\kk$ is 
\[
\begin{split}
\mathcal{Z}_+(I)&=\{P\in\PP(\bar{\kk})^n:\ f(P)=0 \ \mbox{ for all } f\in I\} \\&=\{P\in\PP(\bar{\kk})^n:\ f_1(P)=\ldots=f_r(P)=0\}.
\end{split}
\]
We also denote it by $\mathcal{Z}_+(f_1,\ldots,f_r)$.
\end{definition}

\begin{remark}\label{rem-finitezerolocus}
The following are equivalent for a homogeneous ideal $I\subseteq R$:
$$|\mathcal{Z}(I)|<\infty \Leftrightarrow \mathcal{Z}(I)=\{(0,\dots,0)\} \Leftrightarrow \mathcal{Z}_+(I)=\emptyset.$$
These conditions are equivalent to the fact that the \emph{Krull dimension} of $R/I$ is zero. This is in turn equivalent to $R/I$ being a finite dimensional $k$-vector space.
\end{remark}

\par In Definition~\ref{def-zerodim} and Definition~\ref{def-zerodim-homog} it is important to look at the zero locus of $I$ or $\mathcal{F}$ over the algebraic closure of the base field. 
For cryptographic applications, often the base field $\kk$ is a finite field. In this case the condition that the zero locus is finite over $\kk$ is trivially satisfied by any ideal or system of equations.

\subsection{Infinite fields and the Zariski topology}

Let $\kk$ be a field.
The \emph{Zariski topology} on the affine space $\kk^n$ is the set of complements of solution sets of systems of polynomial equations over $R$, that is $\{\kk^n\setminus\mathcal{Z}(f_1,\ldots,f_r)\mid f_1,\ldots,f_r\in R\}$. If $\kk$ is an algebraically closed field, or at least an infinite field, then every non-empty open set in the Zariski topology is dense, i.e., its closure is equal to the entire space. A non-empty open subset of $\kk^n$ is often called a \emph{generic set} and a property which holds on a non-empty open set is \emph{generic}. Intuitively, a generic set is almost the whole space and a generic property holds almost everywhere in $\kk^n$.

If $\kk$ is a finite field, on the other side, the Zariski topology is the discrete topology on $\kk^n$. In other words, any subset of $\kk^n$ is both open and closed, and the algebraic-geometric intuition of genericity fails. In particular, one can no longer say that a non-empty open subset of $\kk^n$ is almost the whole space, as the closure of any subset of $\kk^n$ is the subset itself. Therefore, as genericity loses its meaning over a finite field, we always will need to assume that the ground field is infinite when dealing with generic sets or properties.

\subsection{Generic changes of coordinates and saturation}

In this paper, we work in the open set defined in \cite{BS87}.
In order to state the definition, we need to recall the algebraic operation of saturation.

\medskip\noindent
{\bf Definition.}
Let $J\subseteq S=R[t]$ be a homogenoeus ideal. The \emph{saturation} of $J$ with respect to the irrelevant maximal ideal of $S$ is $$J^{\sat}=\bigcup_{d\geq 0}\{f\in S\mid fm\in J \mbox{ for every monomial } m \in S_d\}.$$

\begin{definition}\label{gencoord}
Let $\kk$ be an infinite field. Let $J\subseteq S=R[t]$ be a homogeneous ideal with $|\mathcal{Z}_+(J)|<\infty$. We say that $J$ is \emph{in generic coordinates} if either $|\mathcal{Z}_+(J)|=0$ or $t\nmid 0$ mod. $J^{\sat}$.

Let $\kk$ be any field and let $K\supseteq \kk$ with $K$ infinite. $J$ is \emph{in generic coordinates over $K$} if $JK[x_1,\ldots,x_n,t]\subseteq K[x_1,\ldots,x_n,t]$ is in generic coordinates.
\end{definition}

It is easy to see that any homogeneous ideal can be put in generic coordinates by applying a generic change of coordinates to it (see also the proof of \cite[Lemma 2.9]{BS87}). Informally, if $\kk$ is finite, it suffices to apply to $J$ a random change of coordinates over a field extension of sufficiently large cardinality.
\stepcounter{theorem}

\subsection{Homogeneous ideals associated to a system}

\par Let $R=\kk[x_1,\dots,x_n]$ and let $S=R[t]$.
Given a polynomial $f\in R$, we denote by $f^h\in S$ the homogenization of $f$ with respect to the new variable $t$.
For $\mathcal{F}=\{f_1,\dots,f_r\}\subseteq R$, we let $\mathcal{F}^h\subseteq S$ denote the system obtained from 
$\mathcal{F}$ by homogenizing each $f_i$ with respect to $t$, that is $\mathcal{F}^h=\{f_1^h,\dots,f_r^h\}.$

\par For an ideal $I\subseteq R$, the \emph{homogenization} of $I$ with respect to $t$, or simply the homogenization of $I$, is the ideal $$I^{h}=(f^h: \ f\in I)\subseteq S.$$
If $I=(\mathcal{F})\subseteq R$, then $I^h$ is a homogeneous ideal of $S$ which contains $(\mathcal{F}^h)$. It is easy to produce examples where the containment is strict.

\begin{remark}\label{remark-gbasishomogenization}
Let $\mathcal{G}$ be a Gr\"obner basis of $I$ with respect to a degree-compatible term order on $R$. It can be shown that $\mathcal{G}^h=\{g^h: \ g\in\mathcal{G}\}$ is a Gr\"obner basis of $I^h$ with respect to a suitable term order on $S$, see e.g.~\cite[Section 4.3]{KR05}. In particular $I^h=(g^h: \ g\in\mathcal{G})$, hence the degrees of a minimal system of generators of $I^{h}$ are usually different from those of a minimal system of generators of $I$. Instead, the degrees of a minimal system of generators of $(\mathcal{F}^h)$ coincide with the degrees of $f_1,\dots,f_r$.
\end{remark}

\par The \emph{dehomogenization map} $\phi$ is the standard projection on the quotient $\phi:S\rightarrow R\cong S/(t-1)$.
For any system of equations $\mathcal{F}\subseteq R$ generating an ideal $I=(\mathcal{F})$ we have $\phi(I^{h})=(\phi(\mathcal{F}^h))=I$. Notice that one also has $\phi((\mathcal{F}^h))=(\phi(\mathcal{F}^h))=I$.

\par For a polynomial $f\in R$, we denote by $f^{\mathrm{top}}$ its homogeneous part of highest degree.
For a system of equations $\mathcal{F}=\{f_1,\dots,f_r\}$ we denote by 
\begin{equation*}
\mathcal{F}^{\mathrm{top}}=\{f_1^{\mathrm{top}},\dots,f_r^{\mathrm{top}}\}.
\end{equation*}

\par Both the ideal $(\mathcal{F}^h)$ and the ideal $(\mathcal{F}^{\mathrm{top}})$ depend on $\mathcal{F}$, and not only on the ideal $I=(\mathcal{F})$.

\section{The importance of being $LEX$}\label{section-shapelemma}

\par The main goal of this section is clarifying the relation between solving a system of polynomial equations $\mathcal{F}$ and computing a Gr\"obner basis of the ideal $I$ generated by the system. In the cryptographic literature it is often stated that, thanks to the Shape Lemma, the problem of finding the solutions of $\mathcal{F}$ can be reduced to that of computing a lexicographic Gr\"obner basis of $I$. This statement is however not rigorous, since the Shape Lemma only holds under certain assumptions, which are not always verified for cryptographic systems. 
 
We start by stating the assumptions under which the Shape Lemma holds and showing that, when they are satisfied, the problem of solving the system $\mathcal{F}$ is polynomial-time-equivalent to that of computing a lexicographic Gr\"obner basis of $I$. Then we discuss what can be done in the case when the assumptions of the Shape Lemma are not satisfied. We come to the conclusion that, in all situations, one can easily compute the solutions of $\mathcal{F}$ from a lexicographic Gr\"obner basis of $I$. We stress that we are not stating that directly computing the reduced lexicographic Gr\"obner basis is the most efficient way to solve a system (see also Section~\ref{section-solvingdegree}). We conclude the section with a brief discussion of when it is feasible to add the field equations to a system $\mathcal{F}$ and how that affects the computation of a Gr\"obner basis of it.

Throughout the section we focus on systems of equations which have a finite number of solutions over the algebraic closure of the field of definition, since systems that arise in public key cryptography are usually of this kind. Moreover, we always assume that our systems have at least one solution. In fact, if the system has no solutions, the corresponding ideal is equal to the polynomial ring, that is the reduced Gr\"obner basis with respect to any term order is equal to $\{1\}$. In this case, therefore, computing the reduced lexicographic Gr\"obner basis allows us to decide that the system has no solutions, without any additional work.

\medskip

We start by recalling the Shape Lemma.

\begin{theorem}[Shape Lemma -- \cite{KR00}, Theorem 3.7.25]\label{thm:shapelemma}
Let $\kk$ be a field and let $f_1,\dots,f_r\in R$ be such that the corresponding ideal $I=(f_1,\dots,f_r)$ is radical, in normal $x_n$-position, and $|\mathcal{Z}(I)|=d<\infty$. The reduced lexicographic Gr\"obner basis of $I$ is of the form $$\{g_n(x_n),x_{n-1}-g_{n-1}(x_n),\dots,x_1-g_1(x_n)\},$$ where $g_1,\ldots,g_n$ are univariate polynomials in $x_n$ and $\deg(g_1),\ldots,\deg(g_{n-1})<\deg(g_n)=d$. 
\end{theorem}

The Shape Lemma assumes that the ideal $I$ is radical and in normal $x_n$-position.
An ideal $I$ is \emph{radical} if $f^\ell\in I$ for some $\ell>0$ implies $f\in I$. This assumption is not always verified for ideals generated by systems arising in cryptography. Later in the section, we will show how one can use a more general version of the Shape Lemma in order to overcome this problem.

Being in \emph{normal $x_n$-position} means that any two distinct zeros $(a_1,\dots,a_n)$, $(b_1,\dots,b_n)\in\mathcal{Z}(I)$ satisfy $a_n\neq b_n$. 
Notice that every ideal $I$ with finite affine zero locus can be brought into normal $x_n$-position by a suitable linear change of coordinates, passing to a field extension if needed (see \cite[Proposition 3.7.22]{KR00}). A field extension may indeed be needed, as the next example shows.

\begin{example}
Let $\mathcal{F}=\{x_1^2+x_1,x_1x_2,x_2^2+x_2\}\subseteq R=\mathbb{F}_2[x_1,x_2]$. Then $I=(x_1^2+x_1,x_1x_2,x_2^2+x_2)$ is a radical ideal and $\mathcal{Z}(I)=\{(0,0), (0,1), (1,0)\}$. We claim that $I$ cannot be brought in normal $x_2$-position by a linear change of coordinates over $\mathbb{F}_2$. In fact, a linear change of coordinates over $\mathbb{F}_2$ sends $x_2$ to either $x_1$, $x_2$, $x_1+x_2$, $x_1+1$, $x_2+1$, or $x_1+x_2+1$. However, all these linear forms take the same value on at least two of the elements of $\mathcal{Z}(I)$.
\end{example}

Finally, the Shape Lemma assumes that $|\mathcal{Z}(I)|<\infty$. If $\kk$ is a finite field, then one can add the field equations to $I$ and obtain an ideal $J$ which is radical and such that $\mathcal{Z}(J)=\mathcal{Z}(I)\cap\kk^n$, in particular $|\mathcal{Z}(J)|<\infty$. This is however not always advantageous or even feasible, as we discuss in Section~\ref{subs:fieldeqn}. 

Whenever the assumptions of the Shape Lemma are satisfied, computing the solutions of a system of equations has the same complexity as computing the reduced lexicographic Gr\"obner basis of the ideal generated by the system.

\begin{theorem}\label{polytimeequiv}
Let $\mathcal{F}=\{f_1,\dots,f_r\}\subseteq R$ be a polynomial system such that the corresponding ideal $I=(f_1,\dots,f_r)$ is radical and in normal $x_n$-position. Assume that $|\mathcal{Z}(I)|=d<\infty$ and $\mathcal{Z}(I)\subseteq\FF_q^n$. Consider the $LEX$ order.
The set of solutions of $\mathcal{F}$ can be computed from the reduced Gr\"obner basis of $I$ probabilistically in time polynomial in $\log q, n$ and $d$. Conversely, the reduced Gr\"obner basis of $I$ can be computed from the set of solutions of $\mathcal{F}$ deterministically in time polynomial in $\log q, n$ and $d$.
\end{theorem}

\begin{proof}
By the Shape Lemma, the reduced lexicographic Gr\"obner basis of $I$ has the form:
\begin{equation}\label{eqn:redlex}
\{g_n(x_n),x_{n-1}-g_{n-1}(x_n),\dots,x_1-g_1(x_n)\},
\end{equation}
where $g_i(x_n)$ are polynomials in the variable $x_n$ only, and $\deg(g_j)<\deg(g_n)=d$ for $1\leq j<n$.

If we know the reduced lexicographic Gr\"obner basis of $I$, then we can factor the polynomial $g_n(x_n)$ to find its roots. Each root $\alpha$ of $g_n(x_n)$ corresponds to a solution $(g_1(\alpha),\ldots,g_{n-1}(\alpha),\alpha)$ of $f_1=\ldots=f_r=0$.
Notice that the only operation required, apart from the arithmetic over $\FF_q$, is  factoring univariate polynomials, which can be done in probabilistic polynomial time over a finite field.

\par Vice versa, assume that we know $\mathcal{Z}(I)=\{P_1,\dots,P_d\}\subseteq\FF_q^n$ of $\mathcal{F}$. Write $P_i=(a_{i,1},\dots,a_{i,n})$ for $i=1,\dots,d$. We wish to compute the reduced lexicographic Gr\"obner basis of $I$, knowing that it is of the form (\ref{eqn:redlex}).
Since the roots of $g_n$ are exactly $a_{1,n},\dots,a_{d,n}$ we can compute $g_n(x_n)=\prod_{i=1}^d(x_n-a_{i,n})$. 
Now fix $j\in\{1,\dots,n-1\}$. Since $g_j(a_{i,n})=a_{i,j}$ for $i=1,\ldots,d$ and $\deg(g_j)<d$, we can compute $g_j(x_n)$ by using Lagrange interpolation:
\begin{equation*}
g_j(x_n)=\sum_{i=1}^d\left(\prod_{\substack{1\leq\lambda\leq d\\\lambda\neq i}}\frac{x_n-a_{\lambda,n}}{a_{i,n}-a_{\lambda,n}}\right)a_{i,j}.
\end{equation*}
\end{proof}

\medskip
We now discuss the situation in which the assumptions of the Shape Lemma do not hold. In particular, we consider the case when $I$ is not radical. Some authors state that, since $I+(x_1^q-x_1,\ldots,x_n^q-x_n)\subseteq\FF_q[x_1,\ldots,x_n]$ is always radical, up to adding the field equations one may assume without loss of generality that $I$ is radical. However, adding the field equations to the system is not always computationally feasible, even in the case of systems coming from cryptography. Therefore, being able to deal with the situation when the ideal $I$ is not radical is relevant for cryptographic applications. We discuss this issue in more detail in Section~\ref{subs:fieldeqn}. 

Before continuing our discussion, we give an example of system coming from multivariate cryptography for which the corresponding ideal is not radical, adding the field equations to the system is not feasible, and one ends up with a reduced lexicographic Gr\"obner basis which does not have the shape predicted by the Shape Lemma. Indeed, this was the case for most of the instances of the ABC cryptosystem~\cite{TDTD13,TXPD15} that we computed. Since the field sizes proposed in ~\cite{TXPD15} for achieving $80$-bits security are $2^8$, $2^{16}$, and $2^{32}$, adding the field equations to the system is not feasible. 
In our next example we disregard the linear transformations used in the ABC cryptosystem to disguise the private key, since they do not affect the property of the system to generate a radical ideal.

\begin{example}\label{ex-idealwithnoshapebasis}
We consider $R=\FF_2[x_1,x_2,x_3,x_4]$ with the $LEX$ term order and a toy instance of an ABC cryptosystem with
\begin{equation*}
A=\begin{pmatrix}
x_1 & x_2\\
x_3 & x_4 
\end{pmatrix}, \ 
B=\begin{pmatrix}
x_1+x_2+x_3  & x_1+x_2\\
x_1+x_3+x_4 & x_3
\end{pmatrix}, \
C= \begin{pmatrix}
x_1+x_2+x_3+x_4 & x_1+x_4\\
x_1+x_4  & x_1 
\end{pmatrix}.
\end{equation*}
We let $p_1,\dots,p_8$ be the entries of the matrices $AB$ and $AC$. We take a random plaintext $b=(0,1,1,0)\in\mathbb{F}_2^4$ and we evaluate the polynomials $p_1,\dots,p_8$ at $b$ to obtain the ciphertext $a=(1,1,0,1,0,0,0,0)\in\mathbb{F}_2^8$.
We then consider the system $\mathcal{F}=\{p_i-a_i:\ i=1,\dots,8\}$ and the corresponding ideal $I=(\mathcal{F})\subseteq R$. The ideal $I$ is not radical as $(x_3+1)^2 \in I$, but $x_3+1\not\in I$. A computation with MAGMA shows that the reduced lexicographic Gr\"obner basis of $I$ is $\{x_1, x_2+x_3, x_3^2+1, x_4\}$.
\end{example}

We now discuss how one can efficiently compute the solutions of a polynomial system from its lexicographic Gr\"obner basis, without assuming that the ideal generated by the equations is radical. We stress that we always assume that the system has finitely many solutions over the algebraic closure. The next result will be central to our discussion, as we will use it as a substitute of the Shape Lemma.

\begin{theorem}[Elimination Theorem -- \cite{CLO07}, Chapter 3.1, Theorem 2]\label{prop-shapeLEXbasis}
Let $I\subseteq R$ be an ideal and let $\mathcal{G}$ be a lexicographic Gr\"obner basis of $I$. Then for every $1\leq\ell\leq n-1$ the set $\mathcal{G}\cap\kk[x_{\ell+1},\ldots,x_n]$ is a Gr\"obner basis of $I\cap\kk[x_{\ell+1},\ldots,x_n]$ with respect to the $LEX$ order on $\kk[x_{\ell+1},\ldots,x_n]$.
\end{theorem}

In the next result we use Theorem~\ref{prop-shapeLEXbasis} to prove that one can easily compute the solutions of $\mathcal{F}$ from the reduced lexicographic Gr\"obner basis of $I$.

\begin{theorem}\label{thm-nonradsolving}
Let $I$ be a proper ideal of $R=\kk[x_1,\ldots,x_n]$ with finite affine zero locus. The reduced lexicographic Gr\"obner basis of $I$ has the form
\begin{equation*}
\begin{split}
&p_{n,1}(x_n),\\
&p_{n-1,1}(x_{n-1},x_n),\dots,p_{n-1,t_{n-1}}(x_{n-1},x_n),\\
&p_{n-2,1}(x_{n-2},x_{n-1},x_n),\dots,p_{n-2,t_{n-2}}(x_{n-2},x_{n-1},x_n),\\
&\cdots \\
&p_{1,1}(x_1,\dots,x_n),\dots,p_{1,t_1}(x_1,\dots,x_n),
\end{split}
\end{equation*} 
where $p_{i,t_j}\in \kk[x_i,\ldots,x_n]$ for every index $i\in\{1,\dots,n\},j\in\{1,\dots,t_i\}$ and $t_1,\dots,t_{n-1}\geq 1$.
Moreover, for any $1\leq\ell\leq n$, let  $a=(a_{\ell+1},\ldots,a_n)\in\kk^{n-\ell}$ be a solution of the equations
\begin{equation*}
\begin{split}
&p_{n,1}(x_n),\\
&p_{n-1,1}(x_{n-1},x_n),\dots,p_{n-1,t_{n-1}}(x_{n-1},x_n),\\
&\cdots \\
&p_{\ell+1,1}(x_{\ell+1},\dots,x_n),\dots,p_{\ell+1,t_{\ell+1}}(x_{\ell+1},\dots,x_n),
\end{split}
\end{equation*} 
and let $$p_\ell(x_\ell)=\gcd\{p_{\ell,1}(x_\ell,a_{\ell+1},\ldots,a_n),\dots,p_{\ell,t_\ell}(x_\ell,a_{\ell+1},\ldots,a_n)\}.$$
Then $p_\ell(x_\ell)\not\in\kk$.
\end{theorem}

\begin{proof}
Let $\mathcal{G}$ be the reduced lexicographic Gr\"obner basis of $I$. The set $\mathcal{G}\cap\kk[x_{\ell},\ldots,x_n]$ is of the form  $$\mathcal{G}\cap\kk[x_{\ell},\ldots,x_n]=\{p_{i,j}(x_i,\ldots,x_n)\mid \ell\leq i\leq n, 1\leq j\leq t_i\}$$ for some $t_1,\ldots,t_n\geq 0$. 
Moreover, for any $1\leq\ell\leq n$ such that $p_\ell(x_\ell)\neq 0$, one has $t_\ell\geq 1$. Hence it suffices to show that $p_\ell(x_\ell)\not\in\kk$ for $1\leq\ell\leq n$.

We prove the claim by descending induction on $\ell\leq n$. Let $\ell=n$, then $\mathcal{G}\cap\kk[x_n]$ is the reduced lexicographic Gr\"obner basis of $I\cap\kk[x_n]$ by Theorem~\ref{prop-shapeLEXbasis}. Let $p_{n,1}(x_n)$ be a monic generator of $I\cap\kk[x_n]$, then $\mathcal{G}\cap\kk[x_n]=\{p_{n,1}(x_n)\}$ and $t_n=1$. Since the affine zero locus of $I$ is finite, $p_{n,1}(x_n)\neq 0$. Moreover, $p_n(x_n)=p_{n,1}(x_n)\not\in\kk\setminus\{0\}$, since $\emptyset\neq\mathcal{Z}(I)\subseteq\mathcal{Z}(p_{n})$.

We suppose now that the claim holds up to $\ell+1$ and we prove that $p_{\ell}(x_{\ell})\not\in\kk$. By Theorem~\ref{prop-shapeLEXbasis}, $\mathcal{G}\cap\kk[x_{\ell},\ldots,x_n]$ is the reduced lexicographic Gr\"obner basis of $I\cap\kk[x_{\ell},\ldots,x_n]$, in particular $$I\cap\kk[x_{\ell},\ldots,x_n]=(p_{i,j}\mid \ell\leq i\leq n, 1\leq j\leq t_i).$$ 
Let $a\in\mathcal{Z}(I\cap\kk[x_{\ell+1},\ldots,x_n])\cap\kk^{n-\ell}$ and define
$$I(\ell,a)=(p_{\ell,1}(x_\ell,a_{\ell+1},\ldots,a_n),\dots,p_{\ell,t_\ell}(x_\ell,a_{\ell+1},\ldots,a_n))=(p_\ell(x_\ell)).$$
By~\cite[Chapter 3.2, Theorem 3]{CLO07} and since $\mathcal{Z}(I)$ is a finite set, one has that $$\mathcal{Z}(I\cap\kk[x_{\ell},\ldots,x_n])=\pi_{n-\ell+1}(\mathcal{Z}(I)),$$ where $\pi_i:\kk^n\rightarrow\kk^i$ is the projection on the last $i$ coordinates. In particular, $\mathcal{Z}(I\cap\kk[x_{\ell},\ldots,x_n])$ is finite.
If $p_\ell(x_\ell)$ is the zero polynomial, then $\mathcal{Z}(I(\ell,a))=\bar{\kk}$ and 
$$\{(a_\ell,a_{\ell+1},\ldots,a_n)\mid a_\ell\in\bar{\kk}\}\subseteq \mathcal{Z}(I\cap\kk[x_{\ell},\ldots,x_n]),$$ 
contradicting the finiteness of $\mathcal{Z}(I\cap\kk[x_{\ell},\ldots,x_n])$. If instead $p_\ell(x_\ell)\in\kk\setminus\{0\}$, then $\mathcal{Z}(I(\ell,a))=\emptyset$. However, $a=(a_{\ell+1},\ldots,a_n)\in\mathcal{Z}(I\cap\kk[x_{\ell+1},\ldots,x_n])=\pi_{n-\ell}(\mathcal{Z}(I))$, where equality holds by~\cite[Chapter 3.2, Theorem 3]{CLO07}.
So there exist $a_1,\ldots,a_\ell\in\bar{\kk}$ such that $(a_1,\ldots,a_n)\in\mathcal{Z}(I)$. Therefore, $\pi_{n-\ell+1}(a_1,\ldots,a_n)=(a_\ell,\ldots,a_n)\in\mathcal{Z}(I\cap\kk[x_{\ell},\ldots,x_n])$, that is $a_\ell\in\mathcal{Z}(I(\ell,a))=\emptyset$, a contradiction.
\end{proof}

We use the previous result to build an algorithm which computes the affine zero locus of an ideal $I$ from its reduced lexicographic Gr\"obner basis. We adopt the notation of Theorem~\ref{thm-nonradsolving}.

\begin{corollary}\label{cor-algorithmsolutions}
Let $I\subseteq R=\kk[x_1,\dots,x_n]$ be an ideal with finite affine zero locus $\mathcal{Z}(I)$.
Then $\mathcal{Z}(I)$ can be computed as follows:
\begin{enumerate}
\item Compute the reduced lexicographic Gr\"obner basis $\mathcal{G}$ of $I$ to obtain the monic polynomial $p_n\in\kk[x_n]$ such that $(p_n)=I\cap\kk[x_n]$.
\item If $p_n=1$, then $\mathcal{Z}(I)=\emptyset$. Else, factor $p_n$.
\item For every root $\alpha$ of $p_n$ compute $$p_{n-1}(x_{n-1})=\gcd\{p_{n-1,1}(x_{n-1},\alpha),\dots,p_{n-1,t_{n-1}}(x_{n-1},\alpha)\}.$$
\item Factor $p_{n-1}$. 
\item For every root $\beta$ of $p_{n-1}$ compute $$p_{n-2}(x_{n-2})=\gcd\{p_{n-2,1}(x_{n-2},\beta,\alpha),\dots,p_{n-2,t_{n-2}}(x_{n-2},\beta,\alpha)\}.$$
\item Proceed similarly, until all the elements of $\mathcal{Z}(I)$ are found.
\end{enumerate}
\end{corollary}

Notice that the computation is even more efficient under the assumption that the system $\mathcal{F}$, or equivalently the ideal $I$, has only one zero over the algebraic closure.
This is often the case for polynomial systems coming from multivariate cryptosystems, where we usually require that for each ciphertext $b$ there is a unique plaintext $a$ such that $f_i(a)=b$ for every $i=1,\dots r$. 

In such a situation, one does not need to factor any univariate polynomial, since each one of them has exactly one solution, which, for a monic polynomial of degree $d$, can be computed by multiplying the coefficient of $x^{d-1}$ by $-d^{-1}$.

\begin{remark}
Assume that $\kk$ is either a finite field or has characteristic zero.
If $I$ admits only one solution $(a_1,\ldots,a_n)\in\bar{\kk}^n$, then in fact $(a_1,\ldots,a_n)\in\kk^n$. This is true even if the solution has multiplicity higher than one.
In fact, $g_n(x_n)=(x_n-a_n)^d\in\kk[x_n]$, hence $da_n\in\kk$. If $\kk$ has characteristic zero, then $a_n\in\kk$.
Else, let $p$ be the characteristic of $\kk$ and write $d=p^\ell e$ where $p\nmid e$. Then $g_n(x_n)=\left(x_n^{p^\ell}-a_n^{p^\ell}\right)^e\in\kk[x_n]$, so $ea_n^{p^\ell}\in\kk$. This implies $a_n^{p^\ell}\in\kk$, hence $a_n\in\kk$, since $\kk$ is a finite field. One proceeds similarly to prove that $a_i\in\kk$ for all $i$. 
\end{remark}

\begin{remark}
By~\cite[Chapter 3.2, Theorem 3]{CLO07} and since $\mathcal{Z}(I)$ is a finite set, one has that $$\mathcal{Z}(I\cap\kk[x_{\ell},\ldots,x_n])=\pi_{n-\ell+1}(\mathcal{Z}(I))$$ for $1\leq\ell\leq n$, where $\pi_i:\kk^n\rightarrow\kk^i$ is the projection on the last $i$ coordinates. This implies that each path from the roots to the leaves in the tree-shaped computation of Corollary~\ref{cor-algorithmsolutions} produces a solution. In particular, Corollary~\ref{cor-algorithmsolutions} does not perform useless computations.
\end{remark}

\subsection{Adding the field equations to a system}\label{subs:fieldeqn}

Let $\mathcal{Q}=\{x_1^q-x_1,\dots,x_n^q-x_n\}$ be the system consisting of the field equations relative to $\mathbb{F}_q$.
Clearly, for any system of equations $\mathcal{F}=\{f_1,\dots,f_r\}\subseteq R=\FF_q[x_1,\dots,x_n]$ one has $$\mathcal{Z}(\mathcal{F}\cup\mathcal{Q})=\mathcal{Z}(\mathcal{F})\cap\FF_q^n.$$

The systems $\mathcal{F}$ and $\mathcal{F}\cup\mathcal{Q}$, however, often have different algebraic properties. 
It is easy to show that the ideal generated by $\mathcal{F}\cup\mathcal{Q}$ is always radical, while the ideal generated by $\mathcal{F}$ may not be. The structure of the reduced Gr\"obner bases of the ideals generated by the two systems and the degrees of the elements appearing in them are often different as well. As a consequence, adding the field equations to a system often affects the complexity of computing a Gr\"obner basis.

Therefore, passing from $\mathcal{F}$ to $\mathcal{F}\cup\mathcal{Q}$ may or may not provide an advantage. It typically provides an advantage for fields of small size, since the equations of $\mathcal{Q}$ have low degree and adding them to $\mathcal{F}$ makes the ideal radical, a necessary hypothesis for the Shape Lemma (Theorem~\ref{thm:shapelemma}) to apply. 
Over fields of large size, however, adding the field equations may make the computation of a Gr\"obner basis practically infeasible. This is due to the fact that we are adding to the system equations of large degree, which are involved in the computation of a Gr\"obner basis, therefore increasing the degree of the computation. 
In the next example, we show that the solving degree may increase when passing from $\mathcal{F}$ to $\mathcal{F}\cup\mathcal{Q}$ (see Definition \ref{def-solvingdegree} for the definition of solving degree).

\begin{example}\label{ex5}
Let $\mathcal{F}=\{x_3^2-x_2,x_2^3-x_1\}\subseteq\FF_5[x_1,x_2,x_3]$ and let $I=(\mathcal{F})$.
The affine zero locus of $I$ over $\overline{\FF}_5$ is infinite.
If we add the field equations $\mathcal{Q}=\{x_1^5-x_1,x_2^5-x_2,x_3^5-x_3\}$ of $\FF_5$ to $\mathcal{F}$, we obtain the ideal $J=(\mathcal{F}\cup\mathcal{Q})$, which has $\mathcal{Z}(J)=\{(0, 0, 0), (1, 1, 1), (4, 4, 2), (4, 4, 3), (1, 1, 4)\}$. 
The elements of $\mathcal{F}$ are a Gr\"obner basis of $I$ with respect to the LEX order, while the reduced Gr\"obner basis of $J$ with respect to the same order also contains $x_3^5-x_3$. In particular, the Gr\"obner basis of $J$ contains a polynomial of higher degree and one can easily verify that
$$\solvdeg(\mathcal{F}\cup\mathcal{Q})=5>3=\solvdeg(\mathcal{F}).$$ 
\end{example}

Even if we restrict our attention to polynomial systems arising in public-key cryptography, one may not always assume that the field equations can be added to the system. An example coming from multivariate cryptography was given in Example~\ref{ex-idealwithnoshapebasis}.
Another example are systems coming from the relation-collection phase of index calculus on elliptic or hyperelliptic curves, since the field size is very large (e.g., the field size required for $80$-bit security is at least $q\sim 2^{160}$ for an elliptic curve and $q\sim 2^{80}$ for a hyperelliptic curve of genus two).
In such a situation, adding equations of degree $q$ to the system would make it unmanageable.

\section{Solving degree of polynomial systems}\label{section-solvingdegree}

In Section~\ref{section-shapelemma} we discussed how one can compute the solutions of a polynomial system, starting from a lexicographic Gr\"obner basis of the ideal that it generates. In this section, we address the problem of estimating the complexity of computing a lexicographic Gr\"obner basis. In practice, one observes that computing a Gr\"obner basis with respect to $LEX$ is usually slower than with respect to any other term order. On the other hand, computing a Gr\"obner basis with respect to $DRL$ is often faster than with respect to any other term order. Therefore, computing a degree reverse lexicographic Gr\"obner basis and converting it to a lexicographic Gr\"obner basis using FGLM or a similar algorithm is usually more efficient than computing a lexicographic Gr\"obner basis directly. For this reason, in this section we discuss the complexity of computing a Gr\"obner basis of an ideal $I$ in a polynomial ring $R=\kk[x_1,\dots,x_n]$ over a field $\kk$ with respect to the $DRL$ order. We refer the reader to~\cite{FGLM93} for a description of the FGLM algorithm and an estimate of its complexity.

\subsection{Macaulay matrices and solving degree}

We have two main classes of algorithms for computing Gr\"obner bases: \emph{Buchberger's Algorithm} and \emph{linear algebra based algorithms}, which transform the problem of computing a Gr\"obner basis into one or more instances of Gaussian elimination. Examples of linear algebra based algorithms are: $F_4$~\cite{Fau99}, $F_5$~\cite{Fau02}, the \emph{$XL$ Algorithm}~\cite{CKPS00}, and \emph{MutantXL}~\cite{DBMMW08}.
Buchberger's Algorithm is older, and its complexity has been extensively studied. Linear algebra based algorithms are often faster in practice and have contributed to breaking many cryptographic challenges.
However, their complexity is less understood, especially when the input consists of polynomials which are not homogeneous.

In this section, we discuss the complexity of linear algebra based algorithms, which is dominated by Gaussian elimination on the \emph{Macaulay matrices}.
First we describe them for homogeneous systems, following \cite[p. 54]{BFS14}. Let $\mathcal{F}=\{f_1,\dots,f_r\}\subseteq R$ be a system of homogeneous polynomials and fix a term order.
The \emph{homogeneous Macaulay matrix} $M_d$ of $\mathcal{F}$ has columns indexed by the terms of $R_d$ sorted, from left to right, according to the chosen order. The rows of $M_d$ are indexed by the polynomials $m_{i,j}f_j$, where $m_{i,j}\in R$ is a term such that $\deg(m_{i,j}f_j)=d$.
Then the entry $(i,j)$ of $M_d$ is the coefficient of the monomial of column $j$ in the polynomial corresponding to the $i$-th row.

\par Now let $f_1,\dots,f_r$ be any polynomials (not necessarily homogeneous).
For any degree $d\in\mathbb{Z}_+$ the \emph{Macaulay matrix} $M_{\leq d}$ of $\mathcal{F}$ has columns indexed by the terms of $R$ of degree $\leq d$, sorted in decreasing order from left to right. The rows of $M_{\leq d}$ are indexed by the polynomials $m_{i,j}f_j$, where $m_{i,j}$ is a term in $R$ such that $\deg(m_{i,j}f_j)\leq d$. The entries of $M_{\leq d}$ are defined as in the homogeneous case. Notice that, if $f_1,\ldots,f_r$ are homogeneous, the Macaulay matrix $M_{\leq d}$ is just a block matrix, whose blocks are the homogeneous Macaulay matrices $M_d,\ldots,M_0$ associated to the same equations. This is the reason for using homogeneous Macaulay matrices in the case that $f_1,\ldots,f_r$ are homogeneous.

The size of the Macaulay matrices $M_{\leq d}$ and $M_d$, hence the computational complexity of computing their reduced row echelon forms, depends on the degree $d$. Therefore, following \cite{DS13}, we introduce the next definition.

\begin{definition}\label{def-solvingdegree}
Let $\mathcal{F}=\{f_1,\ldots,f_r\}\subseteq R$ and let $\tau$ be a term order on $R$. The \emph{solving degree} of $\mathcal{F}$ is the least degree $d$ such that Gaussian elimination on the Macaulay matrix $M_{\leq d}$ produces a Gr\"obner basis of $\mathcal{F}$ with respect to $\tau$. We denote it by $\solvdeg_{\tau}(\mathcal{F})$. When the term order is clear from the context, we omit the subscript $\tau$.

If $\mathcal{F}$ is homogeneous, we consider the homogeneous Macaulay matrix $M_d$ and let the \emph{solving degree} of $\mathcal{F}$ be the least degree $d$ such that Gaussian elimination on $M_0,\ldots,M_d$ produces a Gr\"obner basis of $\mathcal{F}$ with respect to $\tau$.
\end{definition}

Some algorithms perform Gaussian elimination on the Macaulay matrix for increasing values of $d$. An algorithm of this kind has a termination criterion, which allows to decide whether a Gr\"obner basis has been found and the algorithm can be stopped.
For example, $F_5$ uses the so-called signatures for this purpose.
Other algorithms perform Gaussian elimination on just one Macaulay matrix, for a large enough value of $d$. For such an algorithm, a sharp bound on the solving degree provides a good estimate for the value of $d$ to be chosen. In both cases, the solving degree produces a bound on the complexity of computing the desired Gr\"obner basis. In particular, one may choose to artificially stop a Gr\"obner basis computation in the degree corresponding to the solving degree. For this reason, we use the solving degree to measure the complexity of Gr\"obner bases computations and we do not discuss termination criteria.

\begin{remark}
If $\mathcal{F}$ is not homogeneous, then Gaussian elimination on $M_{\leq d}$ may produce a row that corresponds to a polynomial $f$ such that $\deg(f)<d$ and $\ini(f)$ was not the leading term of any row of $M_{\leq d}$ before performing Gaussian elimination. If this is the case, then some variants of the algorithms add to $M_{\leq d}$ the rows corresponding to the polynomials $mf$, where $m$ is a monomial and $\deg(mf)\leq d$. Then they proceed to compute the reduced row echelon form of this larger matrix. If no Gr\"obner basis is produced in degree $\leq d$, then they proceed by adding to this matrix the appropriate multiples of its rows in the next degree and continue as before. This potentially has the effect of enlarging the span of the rows of $M_{\leq d}$, for all $d$. Introducing this variation may therefore reduce the computational cost of computing a Gr\"obner basis with respect to a given term order, since we might be able to obtain a Gr\"obner basis in a smaller degree than the solving degree, as defined in Definition~\ref{def-solvingdegree}.
Throughout the paper, we consider the situation when \textit{no extra rows are inserted}. Notice that the solving degree is an upper bound on the degree in which the algorithms adopting this variation terminate.
\end{remark}

\begin{definition}
Let $I\subseteq R$ be an ideal and let $\tau$ be a term order on $R$. We denote by $\maxGB_{\tau}(I)$ the maximum degree of a polynomial appearing in the reduced $\tau$ Gr\"obner basis of $I$. If $I=(\mathcal{F})$, we sometimes write $\maxGB_{\tau}(\mathcal{F})$ in place of $\maxGB_{\tau}(I)$.
\end{definition}

It is clear that
\begin{equation*}
\maxGB_{\tau}(\mathcal{F})\leq\solvdeg_{\tau}(\mathcal{F}),
\end{equation*}
for any system of polynomials $\mathcal{F}$ and any degree-compatible term order $\tau$. The inequality may not hold for an arbitrary term order, as we show in the next example. In Example~\ref{ex-solvdegandmaxGB} we show that the inequality may be strict for a degree-compatible term order.

\begin{example}
Let $\mathcal{F}=\{x_3^2-x_2,x_2^3-x_1\}\subseteq\FF_5[x_1,x_2,x_3]$ be the system of Example~\ref{ex5} and let $I=(\mathcal{F})$.
The elements of $\mathcal{F}$ are a Gr\"obner basis of $I$ with respect to the LEX order, while the reduced Gr\"obner basis of $I$ with respect to the same order is $\{x_3^2-x_2,x_3^6-x_1\}$. One can easily verify that
$$\solvdeg_{LEX}(\mathcal{F})=3<6=\maxGB_{LEX}(\mathcal{F}).$$ 

\end{example}

\begin{remark}\label{rem-solvdeg=GBmax}
Assume that $\mathcal{F}=\{f_1,\ldots,f_r\}$ is homogeneous. Gaussian elimination on $M_d$ exclusively produces rows that correspond to polynomials of degree $d$. Therefore
$$\solvdeg_{\tau}(\mathcal{F})=\maxGB_{\tau}(\mathcal{F})$$ for any $\tau$.
\end{remark}

Notice moreover that the solving degree of a system $\mathcal{F}$ may be strictly smaller than the largest degree of an equation of $\mathcal{F}$. This may happen, e.g., when $\mathcal{F}$ contains redundant equations.

\begin{example}
Let $\mathcal{F}=\{x^2+x,xy,y^2+y,x^2y+x^2+x\}\subseteq\mathbb{F}_2[x,y]$. The reduced $DRL$ Gr\"obner basis of $I=(\mathcal{F})$ is $\{x^2+x,xy,y^2+y\}$ and $\solvdeg_{DRL}(\mathcal{F})=2$.
\end{example}

\subsection{Homogenization of ideals and extensions of term order}

We consider a polynomial ring $R=\kk[x_1,\dots,x_n]$ and its extension $S=R[t]$ with respect to a new variable $t$. We compare term orders on $R$ and $S$.

\begin{definition}
Let $\sigma$ be a term order on $R$, let $\tau$ be a term order on $S=R[t]$, and let $\phi:S\rightarrow R$ be the dehomogenization map.
We say that $\tau$ \emph{$\phi$-extends} $\sigma$, or that $\tau$ is a \emph{$\phi$-extension} of $\sigma$, if $\phi(\ini_{\tau}(f))=\ini_{\sigma}(\phi(f))$ for every homogeneous $f\in S$.
\end{definition} 

The next theorem relates Gr\"obner basis and dehomogenization.

\begin{theorem}\label{theorem-dehomog}
Let $\sigma$ be a term order on $R$, and let $\tau$ be a $\phi$-extension of $\sigma$ on $S$.
Let $I$ be an ideal in $R$, let $J$ be a homogeneous ideal in $S$ such that $\phi(J)=I$. The following hold:
\begin{enumerate}
\item $\ini_{\sigma}(I)=\phi(\ini_{\tau}(J))$;
\item if $\{g_1,\dots,g_s\}$ is a homogeneous $\tau$ Gr\"obner basis of $J$, then $\{\phi(g_1),\dots,\phi(g_s)\}$ is a $\sigma$ Gr\"obner basis of $I$.
\end{enumerate}
\end{theorem}
\begin{proof}
We prove \textit{(1)}. Notice that $\ini_{\tau}(J)=(\ini_{\tau}(f): \ f\in J, \ f \text{ homogeneous})$, because $J$ is a homogeneous ideal.
Then we have
\begin{equation*}
\begin{split}
\phi(\ini_{\tau}(J))&=\left(\phi(\ini_{\tau}(f)): \ f\in J, \ f \text{ homogeneous}\right) \\
&=\left(\ini_{\sigma}(\phi(f)): \ f\in J, \ f \text{ homogeneous}\right). \\
\end{split}
\end{equation*}
To conclude the proof of \textit{(1)}, it suffices to show that $$\{\phi(f): \ f\in J, \ f \text{ homogeneous}\}=I.$$
The inclusion from left to right follows from the assumption that $\phi(J)=I$.
To prove the other inclusion, we fix a system of generators $f_1,\dots,f_r$ of $I$ and consider $f=\sum_{i=1}^rp_if_i\in I$, with $p_i\in R$. 
Let $h_i\in J$ be homogeneous such that $\phi(h_i)=f_i$ for all $i$ and define $\tilde{p}=\sum_{i=1}^rt^{\alpha_i}p_i^h h_i$. The polynomial $\tilde{p}$ belongs to $J$ and it is homogeneous for a suitable choice of the $\alpha_i$'s.
Since $\phi(\tilde{p})=\sum_{i=1}^r\phi(t^{\alpha_i}p_i^h h_i)=\sum_{i=1}^rp_if_i=f$, the inclusion follows.

To prove \textit{(2)}, observe that 
\begin{equation*}
\phi(\ini_{\tau}(J))=\left(\phi(\ini_{\tau}(g_i)): \ i=1,\dots,s\right)=\left(\ini_{\sigma}(\phi(g_i))\ i=1,\dots,s\right),
\end{equation*}
since $\phi$ is a homomorphism and $\tau$ $\phi$-extends $\sigma$.
This shows that $\{\phi(g_1),\dots,\phi(g_s)\}$ is a Gr\"obner basis of $\phi(\ini_{\tau}(J))$ with respect to $\sigma$, which is equal to $\ini_{\sigma}(I)$ by \textit{(1)}.
\end{proof}

There is a natural way to $\phi$-extend a term order $\sigma$ on $R$ to a term order $\bar{\sigma}$ on $S$.

\begin{definition}\label{def-barsigma}
Let $m,n$ be terms in $R$, let $\sigma$ be a term order on $R$. Define a term order $\bar{\sigma}$ on $S$ via: $t^{\alpha}m>_{\bar{\sigma}}t^{\beta}n$ if and only if ($m>_{\sigma}n$) or ($m=n$ and $\alpha>\beta$).
\end{definition}

\begin{lemma}\label{lemma-barsigmatermorder}
$\bar{\sigma}$ is a term order on $S$ which $\phi$-extends $\sigma$.
\end{lemma}

\begin{proof}
First we prove that $\bar{\sigma}$ is a term order.
The fact that $1<_{\sigma}m$ for every term $m\in R$ implies $1<_{\bar{\sigma}}m$. We have also $1=t^0<_{\bar{\sigma}}t$.

Now, let $t^{\alpha}m>_{\bar{\sigma}}t^{\beta}n$, with $m,n$ terms in $R$, and $\alpha,\beta\in\mathbb{N}$. We show that $>_{\bar{\sigma}}$ respects multiplication by terms.
We have two possibilities: \textit{1)} $m>_{\sigma}n$ or \textit{2)} $m=n$ and $\alpha>\beta$.
If \textit{1)} holds, then we have $x_im>_{\sigma}x_in$ for every $i=1,\dots,n$ since $\sigma$ is a term order, which implies $x_it^{\alpha}m>_{\bar{\sigma}}x_it^{\beta}n$.
Clearly $t^{\alpha+1}m>_{\bar{\sigma}}t^{\beta+1}n$. If \textit{2)} holds, then $x_im=x_in$ for every $i=1,\dots,n$, therefore $x_it^{\alpha}m>_{\bar{\sigma}}x_it^{\beta}n$ since $\alpha>\beta$. 
Moreover we have $t^{\alpha+1}m>_{\bar{\sigma}}t^{\beta+1}n$, because $m=n$ and $\alpha+1>\beta+1$.

Now we prove that $\bar{\sigma}$ $\phi$-extends $\sigma$, that is $\phi(\ini_{\bar{\sigma}}(f))=\ini_{\sigma}(\phi(f))$ for every $f\in S$ homogeneous. 
Let $f=\sum_{i=1}^da_it^{\alpha_i}m_i$ be a homogeneous polynomial, with $m_i\in R$ distinct terms, $\alpha_i\in\mathbb{N}$, and $a_i\in\kk\setminus\{0\}$.
Then $\phi(f)=\sum_{i=0}^da_im_i$ and $\deg m_i=\deg f-\alpha_i$.
If there is any cancellation in the sum defining $\phi(f)$, then the monomials cancelling have the same degree, then they have already been cancelled in $f$. 
Hence, there is no cancellation in $\phi(f)$.
Without loss of generality, let $m_1=\ini_{\sigma}(\phi(f))$, that is $m_1>_{\sigma}m_i$ for every $i=2,\dots,d$.
Then $t^{\alpha_1}m_1=\ini_{\bar{\sigma}}(f)$, and $\phi(\ini_{\bar{\sigma}}(f))=m_1=\ini_{\sigma}(\phi(f))$.
\end{proof}

\begin{example}
The equality $\phi(\ini_{\bar{\sigma}}(f))=\ini_{\sigma}(\phi(f))$ does not necessarily hold for $f$ not homogeneous.
For example consider $f=tx-x+ty\in S=\kk[x,y,t]$, and let $\sigma=LEX$.
Then $\ini_{\bar{\sigma}}(f)=tx$, $\phi(f)=y$, and $\ini_{\sigma}(\phi(f))=y\neq x=\phi(\ini_{\bar{\sigma}}(f))$.
\end{example}

The next Lemma gives an important example of $\phi$-extension of a term order.

\begin{lemma}\label{lemma-DRLonS}
Fix the $DRL$ order on $R$ and extend it to the $DRL$ order on $S$ by letting $t$ be the smallest variable.
Then the $DRL$ order on $S$ $\phi$-extends the $DRL$ order on $R$.
\end{lemma}

\begin{proof}
\par Let $f=\sum_{i=1}^da_it^{\alpha_i}m_i$ be a homogeneous polynomial, with distinct terms $m_i\in R$, $\alpha_i\in\mathbb{N}$, and $a_i\in\kk\setminus\{0\}$.
Then $\phi(f)=\sum_{i=0}^da_im_i$ and $\deg m_i=\deg f-\alpha_i$.
As in the proof of Lemma \ref{lemma-barsigmatermorder} there is no cancellation in $\phi(f)$.

Without loss of generality, let $\ini_{DRL}(\phi(f))=m_1$, that is $m_1>_{DRL}m_i$ for all $i=2,\dots,d$.
For each $i\in\{2,\dots,d\}$ we have two possibilities: either $\deg m_1>\deg m_i$ or $\deg m_1=\deg m_i$.
If  $\deg m_1>\deg m_i$ then we have $\alpha_1<\alpha_i$, since $\deg m_j+\alpha_j=\deg f$ for every $j$.
This implies $t^{\alpha_1}m_1>_{DRL}t^{\alpha_i}m_i$.
If $\deg m_1=\deg m_i$ then we have $\alpha_1=\alpha_i$, and $t^{\alpha_1}m_1>_{DRL}t^{\alpha_i}m_i$ follows from $m_1>_{DRL}m_i$. Therefore we have $\ini_{DRL}(f)=t^{\alpha_1}m_1$, and $\phi(\ini_{DRL}(f))=m_1=\ini_{DRL}(\phi(f))$.
\end{proof}

\begin{remark}
Fix the $DRL$ order on $R$. The $DRL$ order on $S$ is different from the order $\overline{DRL}$ obtained by applying Definition \ref{def-barsigma}. For example, let $R=\kk[x,y]$ with $x>y$, $S=R[t]$, and consider the monomials $t^3x$ and $ty^2$. We have $t^3x<_{\overline{DRL}}ty^2$ because $x<_{DRL}y^2$ in $R$. In particular, $\overline{DRL}$ is not degree-compatible, while $DRL$ is. Notice however that the two orders coincide on pairs of terms of the same degree.
\end{remark}

\subsection{Solving degree and homogenization}

Let $R=\kk[x_1,\dots,x_n]$ with the $DRL$ order and let $S=R[t]$ with the $DRL$ order with $t$ as smallest variable. Let $\mathcal{F}=\{f_1,\ldots,f_r\}\subseteq R$, let $I=(\mathcal{F})\subseteq R$, let $I^h\subseteq S$ be the homogenization of $I$ with respect to $t$, and let $(\mathcal{F}^h)\subseteq S$ be the ideal generated by $\mathcal{F}^h=\{f_1^h,\ldots,f_r^h\}$. The goal of this section is comparing the solving degree of $\mathcal{F}$, $\mathcal{F}^h$, and $I^h$ with respect to the chosen term orders.
We start with a preliminary result on Gr\"obner bases and homogenization.

\begin{proposition}\label{prop-homogGB}
Let $R=\kk[x_1,\dots,x_n]$ and let $S=R[t]$. Fix the $DRL$ term order on R and extend it to the $DRL$ term order on S by letting t be the smallest variable.
Let $I$ be an ideal of $R$ with Gr\"obner basis $\{g_1,\dots,g_s\}$.
Then $\{g_1^h,\dots,g_s^h\}$ is a Gr\"obner basis of $I^h$.
\end{proposition}

\begin{proof}
First we show that $g_1^h,\dots,g_s^h$ generate $I^h$.
Clearly we have $g_1^h,\dots,g_s^h\in I^h$.
For the other inclusion, consider $f\in I$ of degree $d$ with standard representation $f=\sum_{i=1}^{s}f_ig_i$ for some $f_i\in R$, that is $\ini(f)\geq\ini(f_ig_i)$ for all $i=1,\dots,s$.
\par Since $\ini(f)\geq\ini(f_ig_i)$ and $DRL$ is degree-compatible, we have $d\geq \deg f_i+\deg g_i$.
Therefore we can write
\begin{equation}\label{eq-standardrepresentation}
f^h=\sum_{i=1}^st^{d-\deg f_i -\deg g_i}f_i^hg_i^h,
\end{equation}
which shows that $f^h\in(g_1^h,\dots,g_s^h)$.
\par To prove that $\{g_1^h,\dots,g_s^h\}$ is a Gr\"obner basis, it is enough to show that \eqref{eq-standardrepresentation} is a standard representation for $f^h$, i.e. $\ini(f^h)\geq \ini(t^{d-\deg f_i -\deg g_i}f_i^hg_i^h)$ for all $i=1,\dots,s$.
We observe that $\ini(f^h)=\ini(f)$ does not contain the variable $t$ and we distinguish two cases.
\begin{enumerate}
\item  If $d-\deg f_i -\deg g_i>0$, then a power of $t$ appears in $t^{d-\deg f_i -\deg g_i}f_i^hg_i^h$, and in its initial term as well.
It follows that $\ini(f^h)\geq \ini(t^{d-\deg f_i -\deg g_i}f_i^hg_i^h)$ since  $t$ is the smallest variable in the $DRL$ term order of $S$.
\item If $d-\deg f_i -\deg g_i=0$, then no power of $t$ appears in $\ini(f_i^hg_i^h)$. Therefore we have $\ini(f_i^hg_i^h)=\ini(f_ig_i)\leq\ini(f)=\ini(f^h)$.
\end{enumerate}
\end{proof}

The next result relates the solving degrees of $\mathcal{F}$ and $\mathcal{F}^h$. 

\begin{theorem}\label{cor-solvdegIh}
Let $\mathcal{F}=\{f_1,\ldots,f_r\}\subseteq R=\kk[x_1,\dots,x_n]$ and consider the system $\mathcal{F}^h=\{f_1^h,\ldots,f_r^h\}\subseteq S=R[t]$ obtained from $\mathcal{F}$ by homogenizing $f_1,\ldots,f_r$ with respect to $t$. Let $I^h\subseteq S$ be the homogenization of $I=(\mathcal{F})\subseteq R$ with respect to $t$. Consider the term order $DRL$ on $R$ and $S$, with $t$ as smallest variable.
Then 
\begin{align*}
&\maxGB(\mathcal{F}^h)=\solvdeg(\mathcal{F}^h)\geq\solvdeg(\mathcal{F})\\\geq&\maxGB(\mathcal{F})=\maxGB(I^{h})=\solvdeg(I^{h}).
\end{align*}
\end{theorem}

\begin{proof}
We claim that the Macaulay matrix $M_{\leq d}$ of $\mathcal{F}$ with respect to $DRL$ is equal to the homogeneous Macaulay matrix $M_d$ of $\mathcal{F}^h$ with respect to $DRL$, for every $d\geq1$. 
In fact, the monomials of $S$ of degree $d$ are exactly the homogenizations of the monomials of $R$ of degree $\leq d$.
Similarly, if $m_{i,j}f_j^h$ is the index of a row of $M_d$, i.e., $\deg(m_{i,j}f_j^h)=d$, then $\phi(m_{i,j}f_j^h)=\phi(m_{i,j})f_j$ has degree $\leq d$, hence it is the index of a row of $M_{\leq d}$. 
Conversely, every index $m_{i,j}f_j^h$ of a row of $M_d$, can be obtained from an index of a row of $M_{\leq d}$ by homogenizing and multiplying by an appropriate power of $t$.
In a nutshell, the statement on the columns follows from the fact that $I_{\leq d}=\phi\left((\mathcal{F}^h)_d\right)$. One also needs to check that the order on the columns of $M_d$ and $M_{\leq d}$ is the same.
We consider $M_{\leq d}$. Since $DRL$ is degree-compatible, the columns are ordered in non-increasing degree order from left to right.
The columns of the same degree $j\in\{1,\dots,d\}$ are then ordered according to $DRL$.
Similarly, since $t$ is the smallest variable in the $DRL$ order on $S$, the columns of $M_d$ are ordered in increasing order (from left to right) of powers of $t$, which is equivalent to decreasing order of the degree of the variables $x_1,\dots,x_n$.
Then, the columns with the same power of $t$ are ordered according to $DRL$ on the variables $x_1,\dots,x_n$. This proves that the matrices $M_{\leq d}$ and $M_d$ coincide.

\par Let $I=(\mathcal{F})$ and $J=(\mathcal{F}^h)$. Since the matrices $M_{\leq d}$ and $M_d$ coincide and since the dehomogenization of a Gr\"obner basis of $\mathcal{F}^h$ produces a Gr\"obner basis of $\mathcal{F}$ by Theorem~\ref{theorem-dehomog}, one has $$\solvdeg_{DRL}(\mathcal{F})\leq \solvdeg_{DRL}(\mathcal{F}^h).$$ 

\par The equality $\maxGB(\mathcal{F})=\maxGB(I^h)$ follows from the following two facts:\begin{itemize}
\item By Lemma~\ref{lemma-DRLonS} and Theorem~\ref{theorem-dehomog} the dehomogenization of a DRL Gr\"obner basis of $I^h$ produces a DRL Gr\"obner basis of $I$.
\item The homogenization of a DRL Gr\"obner basis of $I$ produces a DRL Gr\"obner basis of $I^h$ by Proposition~\ref{prop-homogGB}.
\end{itemize}
In particular, no leading term of an element of the reduced Gr\"obner basis of $I^h$ is divisible by $t$, so dehomogenization does not decrease the degrees of the elements of the Gr\"obner basis.
\par Finally, the two equalities 
$$\maxGB(\mathcal{F}^h)=\solvdeg(\mathcal{F}^h) \text{ and }\maxGB(I^{h})=\solvdeg(I^{h})$$ follow from Remark \ref{rem-solvdeg=GBmax}.
\end{proof}

\begin{remark}\label{solvdeg-equality}
Theorem~\ref{cor-solvdegIh} clarifies why, when the system $\mathcal{F}$ is not homogeneous, the largest degree of an element in a reduced Gr\"obner basis may be strictly smaller than the solving degree.
In fact, it is often the case that $\solvdeg_{DRL}(\mathcal{F})=\solvdeg_{DRL}(\mathcal{F}^h)$. In this situation, the difference between $\solvdeg_{DRL}(\mathcal{F})$ and $\maxGB(\mathcal{F})$ is due to the difference between the ideals $(\mathcal{F}^h)$ and $I^h$, and more specifically between $\maxGB(\mathcal{F}^h)$ and $\maxGB(I^h)$.
\end{remark}

The following is an example where $\solvdeg(\mathcal{F})>\maxGB(\mathcal{F})$. See also Example~\ref{ex-tzs} for a cryptographic example.

\begin{example}\label{ex-solvdegandmaxGB}
Let $R=\kk[x,y]$ and let $S=R[t]=\kk[x,y,t]$, both with the $DRL$ order.
We consider the system $\mathcal{F}=\{f_1,f_2\}\subseteq R$ with $f_1=x^2-1$, $f_2=xy+x$, and let $I=(\mathcal{F})$.
Then $\mathcal{F}^h=\{f_1^h,f_2^h\}=\{x^2-t^2,xy+xt\}$, and $I^{h}=(x^2-t^2, y+t)$.
Writing the Macaulay matrices of $\mathcal{F}$, $\mathcal{F}^h$, and $\{x^2-t^2, y+t\}$ and doing Gaussian elimination, one sees that $\solvdeg(\mathcal{F})=\solvdeg(\mathcal{F}^h)=3$.
By computing Gr\"obner bases, one can check that $\maxGB(\mathcal{F}^h)=3$ and $\maxGB(\mathcal{F})=\maxGB(I^h)=2$.
\end{example}

\subsection{Solving degree and Castelnuovo-Mumford regularity}

\par In what follows, we compare the solving degree of a homogeneous ideal with a classical invariant from commutative algebra: the \emph{Castelnuovo-Mumford regularity}.
We recall the definition of this invariant and its basic properties before illustrating the link with the solving degree.

\par Let $R=\kk[x_1,\dots,x_n]$ be a polynomial ring in $n$ variables over a field $\kk$ and let $I$ be a homogeneous ideal of $R$.
For any integer $j$ we recall that $R_j$  denotes the $\kk$-vector space of homogeneous elements of $R$ of degree $j$. 

Choose a minimal system of generators $f_1,\dots,f_{\beta_0}$ of $I$.
We recall that, since $I$ is homogeneous, the number $\beta_0$ and the degrees $d_i=\deg f_i$ are uniquely determined.
We fix an epimorphism $\varphi:R^{\beta_0}\rightarrow I$ sending the canonical basis $\{e_1,\dots,e_{\beta_0}\}$ of the free module $R^{\beta_0}$ to $\{f_1,\dots,f_{\beta_0}\}$.
The map $\varphi$ is in general not homogeneous of degree $0$, so we introduce degree shifts on $R$: For any integer $d$, we denote by $R(-d)$ the $R$-module $R$, whose $j$-th homogeneous component is $R(-d)_j=R_{-d+j}$. 
For example, the variables $x_1,\dots,x_n$ have degree $2$ in $R(-1)$, and degree $0$ in $R(1)$. The map
\begin{equation*}
\varphi:\bigoplus_{j=1}^{\beta_0}R(-d_j)\rightarrow I
\end{equation*}  
is homogeneous of degree $0$, that is $\deg(\varphi(f))=\deg f$ for every $f$.

\par Now consider the submodule  $\ker\varphi\subseteq\bigoplus_{j=1}^{\beta_0}R(-d_j)$.
It is again finitely generated and graded, and is called (first) syzygy module of $I$.
We choose a minimal system of generators of $\ker\varphi$ and we continue similarly defining an epimorphism from a free $R$-module (with appropriate shifts) to $\ker\varphi$ and so on.

\par Hilbert's Syzygy Theorem guarantees that this procedure terminates after a finite number of steps.
Thus, we obtain a \emph{minimal graded free resolution} of $I$:
\begin{equation*}
0\rightarrow F_{p}\rightarrow \cdots \rightarrow F_1\rightarrow F_0\xrightarrow{\varphi} I\rightarrow 0,
\end{equation*}
where the $F_i$ are free $R$-modules of the form
\begin{equation*}
F_i=\bigoplus_{j=0}^{\beta_i}R(-d_{i,j})
\end{equation*}
for appropriate shifts $d_{i,j}\in\mathbb{Z}$.
By regrouping the shifts, we may write the free $R$-modules of the minimal free resolution of $I$ as
\begin{equation*}
F_i=\bigoplus_{j\in\mathbb{Z}}R(-j)^{\beta_{i,j}}.
\end{equation*}
The numbers $\beta_{i,j}=\beta_{i,j}(I)$ are the \emph{(graded) Betti numbers} of $I$.

\begin{definition}\label{def-CMregularity}
The \emph{Castelnuovo-Mumford regularity} of $I$ is
\begin{equation*}
\reg(I)=\max\{j-i:\ \beta_{i,j}(I)\neq 0\}.
\end{equation*}
If $\mathcal{F}$ is a homogeneous system of generators of $I$, we set also $\reg(\mathcal{F})=\reg(I)$.
\end{definition}

\begin{example}\label{ex-regularity}
We consider the ideal $I=(x^2,xy,xz,y^3)$ in $R=\kk[x,y,z]$.
A minimal free resolution of $I$ is given by
\begin{equation*}
0\rightarrow R(-4)\xrightarrow{\varphi_2} R(-3)^3\oplus R(-4)\xrightarrow{\varphi_1} R(-2)^3\oplus R(-3)\xrightarrow{\varphi_0} I\rightarrow0,
\end{equation*}
with $R$-linear maps given by the following matrices
\begin{equation*}
\varphi_0=(x^2, xy, xz, y^3), \ \varphi_1=\begin{pmatrix}
-y & -z & 0 &  0 \\
x & 0 & -z & -y^2 \\
0 & x & y & 0 \\
0 & 0 & 0 & x 
\end{pmatrix}, \ \varphi_2=\begin{pmatrix}
z  \\
-y \\
x \\  
0 
\end{pmatrix}.
\end{equation*}
So the non-zero Betti numbers of $I$ are $\beta_{0,2}=3$, $\beta_{0,3}=1$, $\beta_{1,3}=3$, $\beta_{1,4}=1$, $\beta_{2,4}=1$, and the Castelnuovo-Mumford regularity is $\reg(I)=3$.
\end{example}

For more on regularity and its properties, the interested reader may consult~\cite[Chapter 20]{Eis94} or~\cite{Cha07}.
In the sequel we only mention the facts that are relevant for our purposes.

\begin{remark}
In many texts in commutative algebra or algebraic geometry it is assumed that the field $\kk$ is algebraically closed or infinite.
However, the definition of regularity makes perfect sense over a finite field.
The construction of a minimal free resolution that we illustrated can be carried out over a finite field. Moreover, it shows that the Castelnuovo-Mumford regularity is preserved under field extensions.
In particular, if $I$ is an ideal in a polynomial ring $R=\FF_q[x_1,\dots,x_n]$ over a finite field $\FF_q$ and $J$ is its extension to the polynomial ring $S=\overline{\FF}_q[x_1,\dots,x_n]$ over the algebraic closure of $\FF_q$, then $\reg_R(I)=\reg_S(J)$.
\end{remark}

\par The next theorem is due to Bayer and Stillman. It relates the regularity of a homogeneous ideal to the regularity of its $DRL$ initial ideal. Combined with our Theorem~\ref{cor-solvdegIh}, it will allow us to bound the solving degree of any system.

\begin{theorem}[\cite{BS87}, Theorem~2.4 and Proposition~2.9]\label{theorem-bayerstillman}
Let $J$  be a homogeneous ideal in $\kk[x_1,\dots,x_n]$. Assume that $J$ is in generic coordinates over $\overline{\kk}$, then
\begin{equation*}
\reg(J)=\reg(\ini_{DRL}(J)).
\end{equation*}
\end{theorem}

\begin{remark}\label{rem-reg=maxGB}
If $\kk$ has characteristic zero, after applying a generic change of coordinates to $J$ we have $\reg(\ini_{DRL}(J))=\maxGB_{DRL}(J)$, as shown in~\cite[Proposition 2.9]{BS87}. 
If $\kk$ has positive characteristic, one still has that $$\maxGB_{DRL}(J)\leq\reg(\ini_{DRL}(J))$$ and the inequality is often an equality. In fact this was the case in almost all the examples that we computed while working on this paper. 
Nevertheless, in positive characteristic one can find examples of ideals $J$ in generic coordinates for which the inequality is strict. 
E.g. $J=(x^p,y^p)\subseteq\overline{\FF}_p[x,y]$ is in generic coordinates, $\maxGB_{DRL}(J)=p$, and $\reg(J)=2p-1$.
\end{remark}

\par Combining Theorem~\ref{cor-solvdegIh} and Theorem~\ref{theorem-bayerstillman}, one obtains bounds on the solving degree. Our bounds assume that the ideal generated by the (homogenized) system is in generic coordinates. Notice that this assumption is likely to be satisfied for systems of equations coming from multivariate cryptography, at least over a field of sufficiently large cardinality. In fact, multivariate schemes are often constructed by applying a generic change of coordinates (and a generic linear transformation) to the set of polynomials which constitutes the private key.

For the sake of clarity, we give a homogeneous and a non-homogeneous version of the result. Since the proofs are very similar, and in fact more complicated in the non-homogeneous case, we only give the proof in the latter case.

\begin{theorem}\label{theorem-regularity=solvingdegree-homog}
Let $\mathcal{F}\subseteq R$ be a system of homogeneous polynomials and assume that $(\mathcal{F})$ is in generic coordinates over $\overline{\kk}$. Then
$$\solvdeg_{DRL}(\mathcal{F})\leq\reg(\mathcal{F}).$$
\end{theorem}

The following result allows us to bound the complexity of computing a Gr\"obner basis of a system of equations by establishing a connection with the Castelnuovo-Mumford regularity of the homogenization of the system. 

\begin{theorem}\label{theorem-regularity=solvingdegree}
Let $\mathcal{F}=\{f_1,\ldots,f_r\}\subseteq R$ be a system of polynomials, which is not homogeneous. Let $\mathcal{F}^h=\{f_1^h,\ldots,f_r^h\}\subseteq S=R[t]$ and assume that the ideal $(\mathcal{F}^h)$ is in generic coordinates over $\overline{\kk}$. Then
$$\solvdeg_{DRL}(\mathcal{F})\leq\reg(\mathcal{F}^h).$$
\end{theorem}

\begin{proof}
For a homogeneous ideal $J$ in $R$ or $S$, $\maxGB_{DRL}(J)$ and $\reg(J)$ are invariant under field extension. So we may extend all ideals to the algebraic closure $\overline{\kk}$ of $\kk$.
By Theorem~\ref{cor-solvdegIh} and Theorem \ref{theorem-bayerstillman} we have the chain of equalities and inequalities
 \[
 \begin{split}
 \solvdeg_{DRL}(\mathcal{F})&\leq\solvdeg_{DRL}(\mathcal{F}^h)\\&=\maxGB_{DRL}(\mathcal{F}^h)\leq\reg(\ini_{DRL}(\mathcal{F}^h))=\reg(\mathcal{F}^h).
 \end{split}
 \]
\end{proof}

\begin{remark}
The upper bound in Theorem~\ref{theorem-regularity=solvingdegree-homog} and Theorem~\ref{theorem-regularity=solvingdegree} is often an equality, since generically all the inequalities are in fact equalities if $\kk$ has characteristic zero. This is often the case even if $\kk$ has positive characteristic (see also Remark~\ref{rem-reg=maxGB}).
\end{remark}

By combining Theorem~\ref{theorem-regularity=solvingdegree} and classical results on the Castelnuovo-Mumford regularity (see e.g. \cite[Theorem 12.4]{Cha07}), one immediately obtains the following bound on the solving degree of systems which have finitely many solutions over $\bar{\kk}$. The bound is linear in both the number of variables and the degrees of the polynomials of the system.

\begin{corollary}[Macaulay bound -- \cite{Laz83}, Theorem~2] \label{corollary-solvdegzerodimensional}
Consider a system of equations $\mathcal{F}=\{f_1,\dots,f_r\}\subseteq R$ with $d_i=\deg f_i$ and $d_1\geq d_2\geq\cdots\geq d_r$.
Set $\ell=\min\{n+1,r\}$. Assume that $|\mathcal{Z}_+(\mathcal{F}^h)|<\infty$ and that $(\mathcal{F}^h)$ is in generic coordinates over $\overline{\kk}$. Then 
$$\solvdeg_{DRL}(\mathcal{F})\leq d_1+\ldots+d_\ell-\ell+1.$$
In particular, if $r>n$ and $d=d_1$, then
\begin{equation*}
\solvdeg_{DRL}(\mathcal{F})\leq (n+1)(d-1)+1.
\end{equation*}
\end{corollary}

The condition that $(\mathcal{F}^h)$ is in generic coordinates is not always easy to verify. Nevertheless, if we add the field equations, or their fake Weil descent, to the generators of the ideal, then we can prove that the homogenized system is in generic coordinates.

\begin{theorem}\label{thm-solvdegingencoord}
Let $p>0$ be a prime and let $q=p^e$, $e\geq 1$. Let $\kk$ be a field of characteristic $p$ and let $\mathcal{F}=\{f_1,\dots,f_r\}\subseteq \kk[x_1,\dots,x_n]$ be a system of polynomial equations.
Set $d_i=\deg f_i$ with  $d_1\geq d_2\geq\cdots\geq d_r$ and $\ell=\min\{n+1,r\}$. Assume that one of the following holds:
\begin{enumerate}[(i)]
\item $x_i^q-x_i\in\mathcal{F}$ for $i=1,\ldots,n$, or
\item $x_1^q-x_2,\ldots,x_{n-1}^q-x_n,x_n^q-x_1\in\mathcal{F}$.
\end{enumerate}
Then the ideal $(\mathcal{F}^h)=(f_1^h,\ldots,f_r^h)$ is in generic coordinates over $\bar{\kk}$.
In particular
$$\solvdeg_{DRL}(\mathcal{F})\leq d_1+\ldots+d_\ell-\ell+1.$$
Moreover, if $r>n$ and $d=d_1$, then
\begin{equation*}
\solvdeg_{DRL}(\mathcal{F})\leq (n+1)(d-1)+1.
\end{equation*}
\end{theorem}

\begin{proof}
By definition, $J=(\mathcal{F}^h)$ is in generic coordinates over $\bar{\kk}$ if and only if $t$ is not a zero divisor on the quotient $\bar{\kk}[x_1,\dots,x_n,t]/J^{\sat}$, where $J^{\sat}$ is the saturation of $J$ with respect to the irrelevant maximal ideal $(x_1,\ldots,x_n,t)$. Substituting $t=0$ in the equations of $J$ one obtains the equations $x_1=\ldots=x_n=0$. Therefore the projective zero locus of $J$ does not contain any point with $t=0$. This means that $t\nmid 0$ modulo $J^{\sat}$, hence proving that $J$ is in generic coordinates.
The second part of the statement then follows from Corollary~\ref{corollary-solvdegzerodimensional}.
\end{proof}

\begin{remark}
From the proof of Theorem~\ref{thm-solvdegingencoord} one sees that a system is in generic coordinates whenever it contains equations of the form $x_i^{d_i}+p_i(x_1,\ldots,x_n)$ with $\deg(p_i)<d_i$, for $i=1,\ldots,n$.
\end{remark}

We may use the results established in this section to obtain bounds on the solving degree of the ABC encryption scheme. We assume that the systems have finite affine zero loci, which was the case for all the instances of the ABC cryptosystem  that we computed. 

\begin{example}\label{ex-abc}
The system associated to the ABC cryptosystems \cite{TDTD13,TXPD15} consists of $2n$ quadratic equations in $n$ variables. Therefore by assuming that the system is in generic coordinates, or, if the ground field is $\mathbb{F}_2$, simply by adding the field equations to the system we obtain
$$\solvdeg(\mathcal{F})\leq n+2.$$
\end{example}

\section{Solving degree and degree(s) of regularity}\label{section-dreg}
\par In recent years, different invariants for measuring the complexity of solving a polynomial system of equations were introduced. In particular, the notion of \emph{degree of regularity} gained importance and is widely used nowadays.
In this section we discuss how the degree of regularity is related with the Castelnuovo-Mumford regularity.

\par In the literature we found several definitions of degree of regularity. However, they are mostly variations of the following two concepts:
\begin{enumerate}
\item the degree of regularity by Bardet, Faug\`ere, and Salvy \cite{Bar04,BFS04,BFS14};
\item the degree of regularity by Dubois and Gama, later studied by Ding, Schmidt, and Yang \cite{DG10,DS13,DY13}.
\end{enumerate}
In this section we recall both definitions of degree of regularity and compare them with the Castelnuovo-Mumford regularity.

\subsection{The degree of regularity by Bardet, Faug\`ere, and Salvy}

\par To the best of our knowledge, the degree of regularity appeared first in a paper by Bardet, Faug\`ere, and Salvy~\cite{BFS04} and in Bardet's Ph.D. thesis~\cite{Bar04}.
However, the idea of measuring the complexity of computing the Gr\"obner basis of a homogeneous ideal using its index of regularity can be traced back to Lazard's seminal work~\cite{Laz83}.
Before giving the definition, we recall some concepts from commutative algebra.

\par Let $R=\kk[x_1,\dots,x_n]$ be a polynomial ring over a field $\kk$, let $I$ be a homogeneous ideal of $R$, and let $A=R/I$.
For an integer $d\geq 0$, we recall that $A_d$ denotes the homogeneous part of degree $d$ of $A$.
The function $HF_A(-):\mathbb{N}\rightarrow\mathbb{N}$, $HF_A(d)=\dim_{\kk}A_d$ is called \emph{Hilbert function} of $A$.
It is well known that for large $d$, the Hilbert function of $A$ is a polynomial in $d$ called \emph{Hilbert polynomial} and denoted by $HP_A(d)$. 
The generating series of $HF_A$ is called \emph{Hilbert series} of $A$. We denote it by $HS_A(z)=\sum_{d\in\mathbb{N}}HF_A(d)z^d$.
A classical theorem by Hilbert and Serre says that the Hilbert series of $A$ is a rational function, and more precisely has the form
\begin{equation}\label{eq-hilbertseries}
HS_A(z)=\frac{h_A(z)}{(1-z)^{\ell}}
\end{equation}
where $h_A(z)$ is a polynomial such that $h_A(1)\neq0$, called \emph{h-polynomial} of $A$.

\begin{definition}
The \emph{index of regularity} of $I$ is the smallest integer $\ireg{I}\geq 0$ such that $HF_{R/I}(d)=HP_{R/I}(d)$ for all $d\geq \ireg{I}$. If $\mathcal{F}$ is a system of generators for $I$, we set also $\ireg{\mathcal{F}}=\ireg{I}$.
\end{definition}

\par The index of regularity can be read off the Hilbert series of the ideal, as shown in the next theorem.

\begin{theorem}[\cite{BH98}, Proposition 4.1.12]\label{theorem-iregfromhilbertseries}
Let $I\subseteq R$ be a homogeneous ideal with Hilbert series as in \eqref{eq-hilbertseries} and let $\delta=\deg h_A$. Then $\ireg{I}=\delta-\ell+1$.
\end{theorem}

\par Let $I\subseteq R$ be a homogeneous ideal.
Applying the Grothendieck-Serre's Formula~\cite[Theorem 4.4.3]{BH98} to $R/I$ one obtains 
\begin{equation}\label{eq-indexandregularity}
\ireg{I}\leq\reg(I).
\end{equation}
Moreover, if $I$ is homogeneous and $I_d=R_d$ for $d\gg 0$, then $\ireg{I}=\reg(I)$ by~\cite[Corollary 4.15]{Eis05}. 

\begin{definition}\label{def-dregFaugere}
Let $\mathcal{F}=\{f_1,\dots,f_r\}\subseteq R$ be a system of equations and let $\left(\mathcal{F}^{\mathrm{top}}\right)=(f_1^{\mathrm{top}},\dots,f_r^{\mathrm{top}})$ be the ideal of $R$ generated by the homogeneous part of highest degree of $\mathcal{F}$. Assume that $\left(\mathcal{F}^{\mathrm{top}}\right)_d=R_d$ for $d\gg0$.
The \emph{degree of regularity} of $\mathcal{F}$ is
\begin{equation*}
\dregF{\mathcal{F}}=\ireg{\mathcal{F}^{\mathrm{top}}}.
\end{equation*} 
\end{definition}

\begin{remark}
If $\left(\mathcal{F}^{\mathrm{top}}\right)_d=R_d$ for $d\gg0$, then $|\mathcal{Z}(\mathcal{F})|<\infty$. The converse, however, does not hold in general. See Example~\ref{ex-itopnonzerodim} for an example where $\mathcal{F}$ has finitely many solutions over $\bar{\kk}$, but $\left(\mathcal{F}^{\mathrm{top}}\right)_d\neq R_d$ for all $d$.
\end{remark}

\par The following is an easy consequence of the definitions.

\begin{proposition}\label{prop-dreg&reg}
Let $\mathcal{F}\subseteq R$ be a system of equations. 
Assume that $\left(\mathcal{F}^{\mathrm{top}}\right)_d=R_d$ for $d\gg0$. 
Then $$\dregF{\mathcal{F}}=\reg(\mathcal{F}^{\mathrm{top}}).$$
If in addition $\mathcal{F}$ is homogeneous, then $\mathcal{F}^{\mathrm{top}}=\mathcal{F}$ and $$\dregF{\mathcal{F}}=\reg(\mathcal{F}).$$
\end{proposition}

In the context of multivariate cryptosystems however, it is almost never the case that $\mathcal{F}$ is homogeneous and $\left(\mathcal{F}\right)_d=R_d$ for $d\gg 0$. In fact, this is equivalent to saying that $\mathcal{Z}(I)=\{(0,\ldots,0)\}$ by Remark~\ref{rem-finitezerolocus}. 

\par For a system $\mathcal{F}$ such that $I=(\mathcal{F})$ has finite affine zero locus, we may interpret the condition $\left(\mathcal{F}^{\mathrm{top}}\right)_d=R_d$ for $d\gg0$ as a \textit{genericity} assumption. 
This assumption guarantees that the degree of regularity gives an upper bound on the maximum degree of a polynomial in a Gr\"obner basis of $I$, with respect to any degree-compatible term order.

\begin{remark}
Let $\tau$ be a degree-compatible term order and assume that $\left(\mathcal{F}^{\mathrm{top}}\right)_d=R_d$ for $d\gg0$.  
Let $I=(\mathcal{F})$ and $J=(\mathcal{F}^{\mathrm{top}})$. Then $HP_{R/J}(z)=0$, hence $J_d=\ini_{\tau}(J)_d=R_d$ for $d\geq \dregF{\mathcal{F}}$.
The inclusion $\ini_{\tau}(J)_d\subseteq\ini_{\tau}(I)_d$ holds for any $d$, since $\tau$ is degree-compatible.
So we obtain $\ini_{\tau}(I)_d=R_d$ for $d\geq \dregF{\mathcal{F}}$.
This implies that every element of the reduced Gr\"obner basis of $I$ has degree at most $\dregF{\mathcal{F}}$, that is
\begin{equation}\label{eq-dregbound}
\maxGB_{\tau}(\mathcal{F})\leq \dregF{\mathcal{F}}.
\end{equation}
\end{remark}

\par Notice however that (\ref{eq-dregbound}) does not yield a bound on the solving degree of $\mathcal{F}$, as we show in the next example.

\begin{example}\label{ex-tzs}
We consider the polynomial systems $\mathcal{F}$ obtained in~\cite{BG18} (see also~\cite[Chapter~5]{Bia17}) for collecting relations for index calculus following the approach outlined by Gaudry in~\cite{Gau09}. For $n=3$, they consist of three non-homogeneous equations $f_1,f_2,f_3$ of degree 3 in two variables. Computing 150'000 randomly generated examples of cryptographic size (3 different $q$'s, 5 elliptic curves for each $q$, 10'000 random points per curve), we found that $\left(\mathcal{F}^{\mathrm{top}}\right)_d=R_d$ for $d\gg0$ and $$\solvdeg_{DRL}(\mathcal{F})=\reg(\mathcal{F}^h)=5>4=\dregF{\mathcal{F}}=\ireg{\mathcal{F}^{\mathrm{top}}}.$$ The computations were performed by G. Bianco with MAGMA \cite{BCP97}.
\end{example}

\par Notice moreover that there are systems $\mathcal{F}$ for which $|\mathcal{Z}(\mathcal{F})|<\infty$ and $\left(\mathcal{F}^{\mathrm{top}}\right)_d\neq R_d$ for all $d\geq 0$. 
Definition~\ref{def-dregFaugere} and inequality (\ref{eq-dregbound}) do not apply to such systems. This can happen also for polynomial systems arising in cryptography.

When this happens, one may be tempted to consider $\ireg{\mathcal{F}^{\mathrm{top}}}$ anyway, and use it to bound the solving degree of $\mathcal{F}$. Unfortunately this approach fails since $\ireg{\mathcal{F}^{\mathrm{top}}}$ and $\solvdeg(\mathcal{F})$ might be far apart, as the next examples shows.
On the other hand, the Castelnuovo-Mumford regularity of $\mathcal{F}^h$ still allows us to correctly bound the solving degree of $\mathcal{F}$.

\begin{example}\label{ex-itopnonzerodim}
We consider the polynomial systems obtained in~\cite{GM15} for collecting relations for index calculus following the approach outlined by Gaudry in~\cite{Gau09}. For $n=3$, they consist of three non-homogeneous equations $f_1,f_2,f_3$ in two variables, of degrees 7,7, and 8. Let $\mathcal{F}=\{f_1,f_2,f_3\}$, $\mathcal{F}^h=\{f_1^h,f_2^h,f_3^h\}$, and $\mathcal{F}^{\mathrm{top}}=\{f_1^{\mathrm{top}},f_2^{\mathrm{top}},f_3^{\mathrm{top}}\}$. For 150'000 randomly generated examples of cryptographic size (as in Example~\ref{ex-tzs}) we found that $\solvdeg_{DRL}(\mathcal{F})=\reg(\mathcal{F}^h)=15$, $\left(\mathcal{F}^{\mathrm{top}}\right)_d\neq R_d$ for all $d\geq 0$, and $\ireg{\mathcal{F}^{\mathrm{top}}}=8$. The computations were performed by G. Bianco with MAGMA \cite{BCP97}.
\end{example}

\par Finally, given a polynomial system $\mathcal{F}=\{f_1,\dots,f_r\}$ there is a simple relation between the ideals $(\mathcal{F}^{\mathrm{top}})\subseteq R$ and $(\mathcal{F}^h)\subseteq S$, namely 
\begin{equation}\label{itop&itilde}
(\mathcal{F}^{\mathrm{top}})S+(t)=(\mathcal{F}^h)+(t).
\end{equation} 
Here $(\mathcal{F}^{\mathrm{top}})S$ denotes the extension of $(\mathcal{F}^{\mathrm{top}})$ to $S$, i.e., the ideal of $S$ generated by $\mathcal{F}^{\mathrm{top}}$. Since $\mathcal{F}^{\mathrm{top}}\subseteq R$, $t\nmid 0$ modulo $(\mathcal{F}^{\mathrm{top}})S$. If $t\nmid 0$ modulo $(\mathcal{F}^h)$, then $(\mathcal{F}^h)=(\mathcal{F})^h$ is the homogenization of $(\mathcal{F})$ and 
$\reg(\mathcal{F}^h)=\reg(\mathcal{F}^{\mathrm{top}}).$ Therefore, if $t\nmid 0$ modulo $\mathcal{F}^h$ and $\left(\mathcal{F}^{\mathrm{top}}\right)_d=R_d$ for $d\gg 0$, then $$\dregF{\mathcal{F}}=\reg(\mathcal{F}^h)$$ by Proposition~\ref{prop-dreg&reg}.
However, one expects that in most cases $t\mid 0$ modulo $(\mathcal{F}^h)$. In fact, $(\mathcal{F}^h)=(\mathcal{F})^h$ only in very special cases, namely when $f_1,\ldots,f_r$ are a Macaulay basis of $(\mathcal{F})$ with respect to the standard grading (see~\cite[Theorem~4.3.19]{KR05}). Therefore (\ref{itop&itilde}) usually does not allow us to compare the regularity and the index of regularity of $\mathcal{F}^h$ and $\mathcal{F}^{\mathrm{top}}$. See also~\cite[Section~4.1]{BDDGMT20} for a more detailed discussion.

\subsection{The degree of regularity by Ding and Schmidt}

\par The second notion of degree of regularity is more recent. 
To the extent of our knowledge it has been introduced by Dubois and Gama \cite{DG10}, and later has been used by several authors such as Ding, Schmidt, and Yang \cite{DS13,DY13}.
The definition we present here is taken from \cite{DS13}, and differs slightly from the original one of Dubois and Gama.

\par Let $\FF_q$ and let $B=\FF_q[x_1,\dots,x_n]/(x_1^q,\dots,x_n^q)$. 
Let $f_1,\dots,f_r\in B$ be homogeneous polynomials of degree $2$.
We fix a $B$-module homomorphism $\varphi$ sending the canonical basis $e_1,\dots,e_r$ of $B^r$ to $\{f_1,\dots,f_r\}$, that is for every $(b_1,\dots,b_r)\in B^r$ we have $\varphi(b_1,\dots,b_r)=\sum_{i=1}^rb_if_i$.
We denote by $\Syz(f_1,\dots,f_r)$ the first syzygy module of $f_1,\dots,f_r$, that is the kernel of $\varphi$. 
An element of $\Syz(f_1,\dots,f_r)$ is a syzygy of $f_1,\dots,f_r$. In other words, it is a vector of polynomials $(b_1,\dots,b_r)\in B^r$ such that $\sum_{i=1}^rb_if_i=0$.

\par An example of syzygy is given by the Koszul syzygies $f_ie_j-f_je_i$, where  $i\neq j$ or by the syzygies coming by the quotient structure of $B$, that is $f_i^{q-1}e_i$. Here $e_i$ denotes the $i$-th element of the canonical basis of $B$. These syzygies are called \emph{trivial syzygies}, because they are always present and do not depend on the structure of $f_1,\dots,f_r$, but rather on the ring structure of $B$.
We define the module $\Triv(f_1,\dots,f_r)$ of trivial syzygies of $f_1,\dots,f_r$ as the submodule of $\Syz(f_1,\dots,f_r)$ generated by $\{ f_ie_j-f_je_i: \ 1\leq i<j\leq r \}\cup\{f_i^{q-1}e_i: \ 1\leq i\leq r \}$.

For any $d\in\mathbb{N}$ we define the vector space $\Syz(\mathcal{F})_d=\Syz(\mathcal{F})\cap B_d^r$ of syzygies of degree $d$.
We define the vector subspace of trivial syzygies of degree $d$ as $\Triv(\mathcal{F})_d=\Triv(\mathcal{F})\cap B_d^r$.
Clearly, we have $\Triv(\mathcal{F})_d\subseteq\Syz(\mathcal{F})_d$.

\begin{definition}\label{def-dregDing}
Let $\mathcal{F}=\{f_1,\dots,f_r\}\subseteq B$ be a system of polynomials of degree $2$.
The \emph{degree of regularity} of $\mathcal{F}$ is 
\begin{equation*}
\dregD{\mathcal{F}}=\min\{d\geq 2: \ \Syz(\mathcal{F}^{\mathrm{top}})_{d-2}/\Triv(\mathcal{F}^{\mathrm{top}})_{d-2}\neq0 \}.
\end{equation*}
\end{definition}

\begin{remark}
Dubois and Gama \cite{DG10} work in the ring $\FF_q[x_1,\dots,x_n]/(x_1^q-x_1,\dots,x_n^q-x_n)$ and not in $B=\FF_q[x_1,\dots,x_n]/(x_1^q,\dots,x_n^q)$.
\end{remark}

\par The degree of regularity is the first degree where we have a linear combination of multiples of $f_1,\dots,f_r$ which produces a non-trivial cancellation of their top degree parts.
For this reason, some authors refer to it as \emph{first fall degree}.

\par One may wonder whether the degree of regularity by Ding and Schmidt is close to the solving degree of a polynomial system of quadratic equations.
Ding and Schmidt showed that this is not always the case.
In fact, it is easy to produce examples, the so-called degenerate systems, for which the degree of regularity and the solving degree are far apart.
For a detailed exposition on this problem and several examples we refer the reader to their paper \cite{DS13}.

\par We are not aware of any results relating $\dregD{\mathcal{F}}$ (Definition \ref{def-dregDing}) and $\dregF{\mathcal{F}}$ (Definition \ref{def-dregFaugere}).
Despite the fact that they share the name, we do not see an immediate connection. A comparison between these two invariants is beyond the scope of this paper.

\section{Solving degree of ideals of minors and the MinRank Problem}\label{section-minrank}

The goal of this section is giving an example of how the results from Section~\ref{section-solvingdegree}, 
in combination with known commutative algebra results, allow us to prove estimates for the solving degree in a simple and synthetic way.
We consider polynomial systems coming from the MinRank Problem. For more bounds on the complexity of the MinRank Problem, see~\cite{CG19}.

\par The MinRank Problem can be stated as follows.
Given an integer $t\geq1$ and a set $\{M_1,\dots,M_n\}$ of $s\times s$ matrices with entries in a field $\kk$, find a non-zero tuple $\lambda=(\lambda_1,\dots,\lambda_n)\in\kk^n$ such that 
\begin{equation}\label{eq-minrank}
\mathrm{rank}\left(\sum_{i=1}^n\lambda_iM_i\right)\leq t-1.
\end{equation}
This problem finds several applications in multivariate cryptography and in other areas of cryptography as well.
For example, Goubin and Courtois \cite{GC00} solved a MinRank Problem to attack Stepwise Triangular Systems, and Kipnis and Shamir \cite{KS99} solved an instance of MinRank in their cryptanalysis of the HFE cryptosystem.  

\par Consider the matrix $M=\sum_{i=1}^n x_iM_i$, whose entries are homogeneous linear forms in $R$.
Condition~\eqref{eq-minrank} is equivalent to requiring that the minors of size $t\times t$ of $M$ vanish. 
Therefore, every solution of the MinRank Problem corresponds to a non-zero point in the zero locus in $\kk^n$ of the ideal $I_t(M)$ of $t$-minors of $M$.
A similar algebraic formulation can be given for the Generalized MinRank Problem, which finds applications within coding theory, non-linear computational geometry, real geometry, and optimization. We refer the interested reader to~\cite{FSS13} for a discussion of the applications of the Generalized MinRank Problem and a list of references.

\begin{problem}[Generalized MinRank Problem] 
Given a field $\kk$, an $r\times s$ matrix $M$ whose entries are polynomials in $R=k[x_1,\ldots,x_n]$, and an integer $1\leq t\leq\min\{r,s\}$, find a point in $\kk^n\setminus\{(0,\ldots,0)\}$ at which the evaluation of $M$ has rank at most $t-1$. 
\end{problem}

\par The Generalized MinRank Problem can be solved by computing the zero locus of the ideal of $t$-minors $I_t(M)$. The minors of size $t\times t$ of the matrix $M$ form an algebraic system of multivariate polynomials, which one can attempt to solve by computing a Gr\"obner basis.
This motivates our interest in estimating the solving degree of this system for large classes of matrices.

\par Ideals of minors of a matrix with entries in a polynomial ring are called \emph{determinantal ideals} and have been extensively studied in commutative algebra and algebraic geometry.
Using Theorem \ref{theorem-regularity=solvingdegree-homog}, we can take advantage of the literature on the regularity of determinantal ideals to give bounds on the solving degree of systems of minors of certain large classes of matrices. For simplicity, we focus on homogeneous matrices.

\begin{definition}
Let $M$ be an $r\times s$ matrix with $r\leq s$, whose entries are elements of $R$. The matrix $M$ is {\em homogeneous} if both its entries and its $2$-minors are homogeneous polynomials. 
\end{definition}

\par It is easy to see that the minors of any size of a homogeneous matrix are homogeneous polynomials. Moreover, observe that a matrix whose entries are homogeneous polynomials of the same degree is a homogeneous matrix, but there are homogeneous matrices whose entries have different degrees.
After possibly exchanging some rows and columns, we may assume without loss of generality that the degrees of the entries of a homogeneous matrix increase from left to right and from top to bottom. With this notation, we can compute the solving degree of our first family of systems of minors. We refer the reader to~\cite{Eis94} for the definition of height of an ideal.

\begin{theorem}\label{thm-en}
Let $M=(f_{ij})$ be an $r\times s$ homogeneous matrix with $r\leq s$, whose entries are elements of $R$, $n\geq s-r+1$.  
Let $\mathcal{F}$ be the polynomial system of the minors of size $r$ of $M$. Assume that $\height (I_r(M))=s-r+1$ and that $I_r(M)$ is in generic coordinates. Then the solving degree of $\mathcal{F}$ is upper bounded by
$$\solvdeg(\mathcal{F})\leq \deg(f_{1,1})+\ldots+\deg(f_{m,m})+\deg(f_{m,m+1})+\ldots+\deg(f_{m,n})-s+r.$$
If $\deg(f_{i,j})=1$ for all $i,j$, then $\solvdeg(\mathcal{F})=r$.
\end{theorem}

\begin{proof}
Since the matrix $M$ is homogeneous, the system of minors $\mathcal{F}$ consists of homogeneous polynomials. The regularity of the corresponding ideal $I_r(M)=(\mathcal{F})$ is 
$$\reg(I_r(M))=\deg(f_{1,1})+\ldots+\deg(f_{r,r})+\deg(f_{r,r+1})+\ldots+\deg(f_{r,s})-s+r.$$ 
The formula can be found in~\cite[Proposition~2.4]{BCG04} and is derived from a classical result of Eagon and Northcott \cite{EN62}.
The bound on the solving degree now follows from Theorem~\ref{theorem-regularity=solvingdegree-homog}.
In particular, if $\deg(f_{i,j})=1$ for all $i,j$, then $\solvdeg (\mathcal{F})\leq r$. Since $I_r(M)$ is generated in degree $r$, then $\solvdeg (\mathcal{F})=r$.
\end{proof}

Notice that the assumption on the height is satisfied by a matrix $M$ whose entries are generic homogeneous polynomials of fixed degrees. 
If $n=s-r+1$, then $I_r(M)_d=R_d$ for $d\gg 0$, hence $\dregF{\mathcal{F}}=\reg(\mathcal{F})$, where $\mathcal{F}$ is the set of maximal minors of $M$. 
Therefore, Theorem~\ref{thm-en} recovers the results of~\cite{FSS10,FSS13} for $n=s-r+1$ and $t=r$, 
and extends them to homogeneous matrices whose entries do not necessarily have the same degree.
\medskip 

We now restrict to systems of maximal minors of matrices of linear forms.
The MinRank Problem associated to this class of matrices is a slight generalization of the classical MinRank Problem of \eqref{eq-minrank}. 
From the previous result it follows that, if the height of the ideal of maximal minors is as large as possible, then the solving degree of the corresponding system is as small as possible, namely $r$. 
We now give different assumptions which allows us to obtain the same estimate on the solving degree, for ideals of maximal minors whose height is not maximal. 
We are also able to bound the solving degree of the system of $2$-minors.

Let $R$ have a standard $\mathbb{Z}^v$-graded structure, i.e., the degree of every indeterminate of $R$ is an element of the canonical basis $\{e_1,\dots,e_v\}$ of $\mathbb{Z}^v$.

\begin{definition}
Let $M=(f_{i,j})$ be an $r\times s$ matrix with entries in $R$, $r\leq s$.
We say that $M$ is \emph{column-graded} if $s\leq v$, and $f_{i,j}=0$ or it is homogeneous of degree $\deg(f_{i,j})=e_j\in\mathbb{Z}^v$ for every $i,j$.
We say that $M$ is \emph{row-graded} if $r\leq v$, and $f_{i,j}=0$ or it is homogeneous of degree $\deg(f_{i,j})=e_i\in\mathbb{Z}^v$ for every $i,j$. 
\end{definition}
Informally, a matrix is row-graded if the entries of each row are homogeneous linear forms in a different set of variables. Similarly for a column-graded matrix.

\begin{theorem}\label{CSideals}
Let $r\leq s$ and let $M$ be an $r\times s$ row-graded or column-graded matrix with entries in $R$ . Assume that $I_r(M)\neq0$ and that $I_r(M)$ is in generic coordinates.
Then:
\begin{itemize}
\item if $\mathcal{F}$ is the system of maximal minors of $M$ then $\solvdeg(\mathcal{F})=r$,
\item if $\mathcal{F}$ is the system of $2$-minors of $M$ then $\solvdeg(\mathcal{F})\leq s$ in the column-graded case, and $\solvdeg(\mathcal{F})\leq r$ in the row-graded case.
\end{itemize}
\end{theorem}

\begin{proof}
It is shown in~\cite{CDG15,CDG16} that $\reg(I_r(M))=r$, $\reg(I_2(M))\leq s$ in the column-graded case, and $\reg(I_2(M))\leq r$ in the row-graded case. 
The bounds on the solving degree now follow from Theorem~\ref{theorem-regularity=solvingdegree-homog}.
\end{proof}

%
%
%
%

\end{document}